\documentclass[journal]{IEEEtran}

\usepackage{cite}

\usepackage{amsthm}
\usepackage{amsmath}
\usepackage{amsfonts}

\usepackage{hyperref}
\usepackage{microtype}
\usepackage{enumerate}
\usepackage{booktabs}
\usepackage{graphicx}
\usepackage{xcolor}
\usepackage{xspace}

\hypersetup{
  allcolors = black,
  colorlinks = true
}

\newtheoremstyle{examplestyle}
{0.4em}{0.4em}{\normalfont}{}{\bfseries}{.}{.5em}{}
\theoremstyle{examplestyle}\newtheorem{example}{Example}[section] 

\newtheoremstyle{definitionstyle}
{0.4em}{0.4em}{\itshape}{}{\bfseries}{.}{.5em}{}
\theoremstyle{definitionstyle}\newtheorem{definition}{Definition}[section] 

\newcommand{\metname}[0]{{\small\textup{PAGURUS}}\xspace}

\renewcommand{\tablename}{Table}
\newcommand{\fillparagraph}{\unskip\parfillskip 0pt \par}
\newcommand{\acc}[1]{{\footnotesize\textup{#1}}\xspace}

\begin{document}
%
% Title
%
\title{PAGURUS: Low-Overhead Dynamic Information \\ 
Flow Tracking on Loosely Coupled Accelerators}
%
% Authors
%
\author{Luca~Piccolboni,~\IEEEmembership{Student Member,~IEEE,}
        Giuseppe~Di~Guglielmo,~\IEEEmembership{Member,~IEEE,}
        and~Luca~P.~Carloni,~\IEEEmembership{Fellow,~IEEE}
%
% Thanks
%
\thanks{The authors are within the Department of Computer Science, Columbia
University, New York, 10027, NY, USA. Emails: piccolboni@cs.columbia.edu,
giuseppe@cs.columbia.edu, luca@cs.columbia.edu. This article was presented in
the International Conference on Hardware/Software Codesign and System Synthesis
(CODES+ISSS) 2018 and appears as part of the ESWEEK-TCAD special issue.
Accepted July 2, 2018. DOI: 10.1109/TCAD.2018.2857321}}

% Paper headers

\markboth{IEEE TRANSACTIONS ON COMPUTER-AIDED DESIGN OF INTEGRATED CIRCUITS AND
SYSTEMS}{Piccolboni \MakeLowercase{\textit{et al.}}: IEEE TRANSACTIONS
ON COMPUTER-AIDED DESIGN OF INTEGRATED CIRCUITS AND SYSTEMS}%

\maketitle

%
% Abstract
%

% ==> increase line distance
\renewcommand{\baselinestretch}{1.08}

\begin{abstract}%
Software-based attacks exploit bugs or vulnerabilities to get unauthorized
access or leak confidential information.  Dynamic information flow tracking
({\footnotesize DIFT}) is a security technique to track spurious information
flows and provide strong security guarantees against such attacks. To secure
{heterogeneous systems}, the spurious information flows must be tracked through
all their components, including processors, accelerators (i.e.,
application-specific hardware components) and memories. We present
{\footnotesize PAGURUS}, a flexible methodology to design a low-overhead shell
circuit that adds {\footnotesize DIFT} support to accelerators.  The shell uses
a coarse-grain {\footnotesize DIFT} approach, thus not requiring to make
modifications to the accelerator's implementation. We analyze the performance
and area overhead of the {\footnotesize DIFT} shell on {\footnotesize FPGAs}
and we propose a metric, called information leakage, to measure its {security
guarantees}.  We perform a design-space exploration to show that we can
synthesize accelerators with different characteristics in terms of performance,
cost and security guarantees.  We also present a case study where we use the
{\footnotesize DIFT} shell to secure an accelerator running on a embedded
platform with a {\footnotesize DIFT}-enhanced {\footnotesize RISC-V} core.
\fillparagraph
\end{abstract}

% ==> increase line distance
\renewcommand{\baselinestretch}{1.08}


% ==> original line distance
\renewcommand{\baselinestretch}{1.0}

\IEEEpeerreviewmaketitle

%
% Introduction
%

%
\vspace{-0.6cm}%
\section{Introduction}
\label{section:introduction}

{
\IEEEPARstart{H}{eterogeneous} systems-on-chip (SoCs) include multiple
processor cores and application-specific hardware components, known as hardware
\textit{accelerators}, to reduce power consumption and increase
performance~\cite{carloni2016, horowitz2014, cong2014, khailany2018}. Several
accelerators and accelerator-rich architectures have been developed for
different applications, including neural networks~\cite{reagen2016, chen2017},
database processing~\cite{sukhwani2014, wu2015}, graph
processing~\cite{ahn2015, ham2016}, and biomedical
applications~\cite{pagliari2017}.
There exist two main models of accelerators~\cite{cota2015}. Tightly coupled
accelerators are embedded within the processor cores as application-specific
functional units~\cite{srinivasan2011}. They are well-suited for fine-grain
computations on small data sets.  They require to extend the instruction set
architecture of the processor cores to include special instructions and manage
their execution. Loosely coupled accelerators, instead, reside outside the
processor cores. They typically achieve high speed-ups with coarse-grain
computations on big data sets~\cite{piccolboni2017a}. They are called by
software applications through device drivers.
\fillparagraph
}

\smallskip

{
Software-based attacks can exploit security vulnerabilities or bugs in software
applications, e.g., buffer overflows and format strings, to obtain unauthorized
control of applications, inject malicious code, etc.~\cite{whitman2003}.
\textit{Dynamic information flow tracking} (DIFT), also known as {dynamic taint
analysis} in the literature, has been proposed as a promising security
technique to protect systems against software attacks~\cite{suh2004,
newsome2005}.
DIFT is based on the observations that (1) it is impossible to prevent the
injection of untrustworthy data in software applications (e.g., data coming
from software users), and (2) it is very difficult to cover all the possible
exploits that use such data. It is better to monitor, i.e., {track}, the
suspicious data flows during the application execution to ensure that they are
not exploited and do not cause a security violation. In such a protection
scheme, the data flows from the untrustworthy sources are marked as spurious. A
security policy imposes what the system is allowed to do with spurious data.
For example, a policy can enforce that spurious data values are never used as
pointers, thus avoiding buffer-overflow attacks.
\fillparagraph
}

\smallskip

{ 
Several implementations of DIFT have been proposed in the literature. DIFT has
been implemented in hardware~\cite{suh2004, crandall2004, venka2008,
palmiero2018} as well as software~\cite{qin2006, clause2007}. DIFT has been
shown to be effective in protecting systems against several software-based
attacks, including leakage of information~\cite{enck2014} and code
injection~\cite{ardesh2017}.  DIFT is now implemented on different types of
architectures~\cite{dalton2007, venka2008}, including smartphones~\cite{gu2013,
enck2014}. Most of the approaches on hardware-based DIFT focused only on
securing processor cores and the associated logic, i.e., tightly coupled
accelerators, memories and communication channels. Loosely coupled
accelerators, however, have been shown to be vulnerable to
attacks~\cite{olson2016, pilato2017} and to date there have been only two
works~\cite{porquet2013, pilato2018} on DIFT considering such accelerators. We
propose \metname as a methodology to extend the support of DIFT to
loosely coupled accelerators in heterogeneous SoCs.
\fillparagraph 
}
 
\smallskip

%
% Contributions
%

\textbf{Contributions.} We make the following contributions:

\begin{enumerate}

{
\item[\textbf{(1)}]
we present \metname, a flexible methodology to design a low-overhead DIFT shell
that secures loosely coupled accelerators; a shell is a hardware circuit whose
design is \textit{independent} from the design of the accelerators, thus
simplifying the integration of DIFT in heterogeneous SoCs; we analyze the
performance and cost overhead of the shell by synthesizing and running it on
FPGAs: the shell has a low impact on execution time and area of the
accelerators; 
\fillparagraph
}

{
\item[\textbf{(2)}] 
we define the metric of \textit{information leakage} for accelerators to
quantitatively measure the security of the DIFT shell: we show that, for any
given accelerator, it is possible to find the minimum number of tags (required
by DIFT) so that no information leakage is possible; we also show that few tags
interleaved in the accelerators data are often sufficient to guarantee the
absence of information leakage;
\fillparagraph
}

{
\item[\textbf{(3)}]
we perform a design-space exploration where we consider performance, cost and
information leakage as optimization goals for the accelerators design: this
study shows how to strengthen the security of hardware-accelerated applications
in exchange of lower performance and higher cost;
\fillparagraph
}

{
\item[\textbf{(4)}]
we present a case study where the DIFT shell has been used to protect an
accelerator integrated on a embedded SoC~\cite{gautschi2017} we extended with
DIFT: this shows why a holistic DIFT approach is necessary for heterogeneous
SoCs.
\fillparagraph
}

\end{enumerate}

%
% Preliminaries 
%

\begin{figure}[t]
    \centering
    \includegraphics[width=0.96\linewidth]{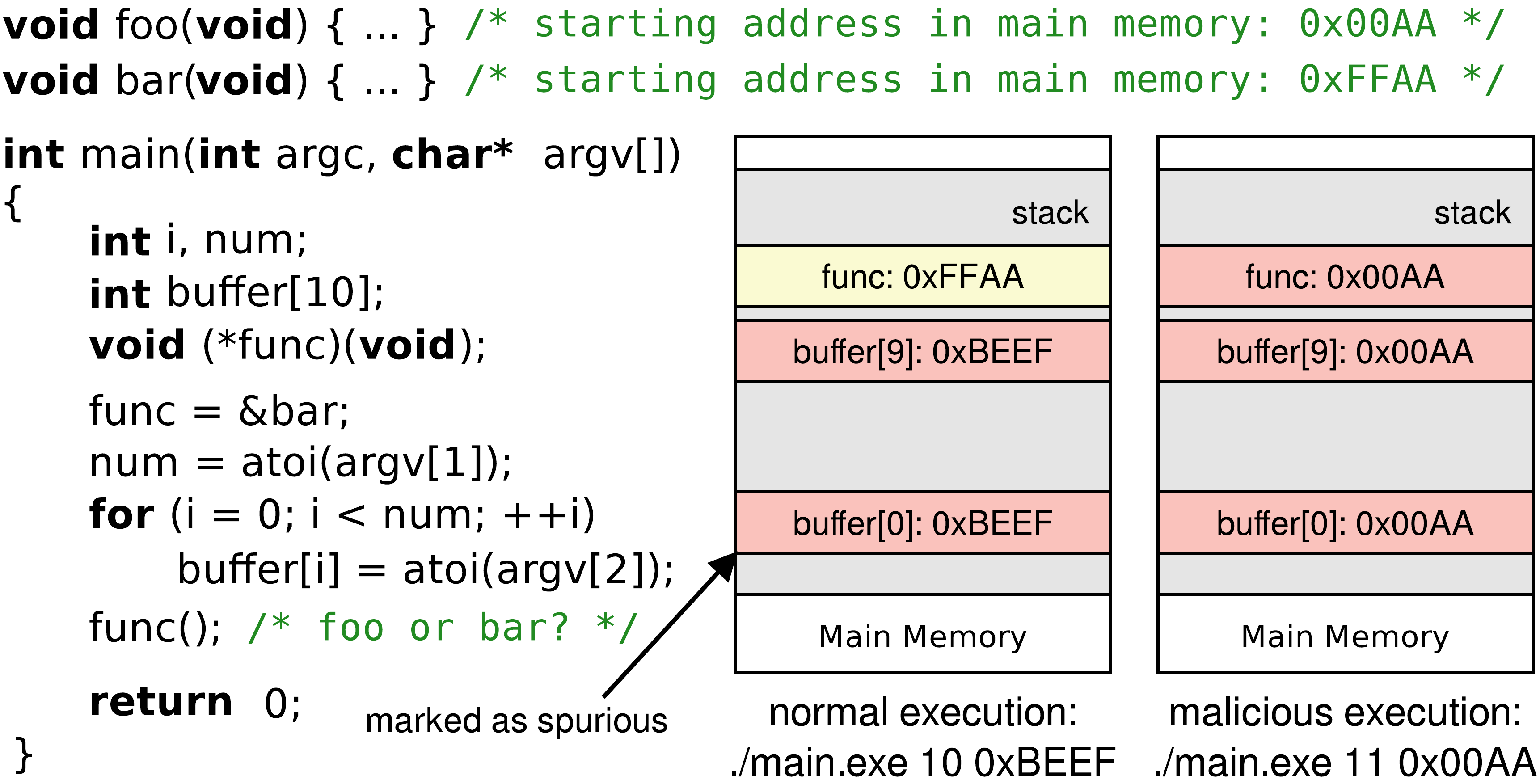}
    \caption{Example of a simple stack-based buffer-overflow attack.}
    \label{figure:introattack}
\end{figure}

\section{Preliminaries}
\label{section:prelim}

{
This section provides the background. We first describe how DIFT prevents a
buffer-overflow attack in practice. Then, we present the architecture of the
SoCs and accelerators we target.  Finally, we discuss our assumptions and the
attack model.
}

%
% Dynamic Information Flow Tracking (DIFT)
%

%
\subsection{Dynamic Information Flow Tracking (DIFT)}
\label{section:prelim:dift}

{
DIFT is a security technique implemented either in hardware or software to
prevent software-based attacks~\cite{suh2004}, e.g., buffer overflows. It has
been also used, for instance, to avoid leakage of information~\cite{enck2014}
and secure Web applications~\cite{lam2008}. The key idea is to use
\textit{tags} to mark as spurious the data generated by untrustworthy channels,
e.g., the input provided by the user to the application. DIFT decouples the
concepts of \textit{policy} (what to do) and \textit{mechanism} (how to do it).
The security {policy} defines which are the untrustworthy channels and the
restrictions to apply on using the data marked as spurious. The {mechanism}
ensures that the untrustworthy data are marked as spurious and the tags are
propagated in the rest of the system. The presence of tags is transparent to
both software users and programmers.
}

\begin{example}
Consider the stack-based buffer-overflow attack of
\figurename~\ref{figure:introattack}.  While there are other ways to prevent
such attack, e.g., non-executable stack, this simple example illustrates a
possible application of DIFT.  If the user specifies a number of iterations
{\small\texttt{num}} higher than $10$, then the function pointer
{\small\texttt{func}} can be overwritten. In this case, another function
({\small\texttt{foo}}) can be executed (see the stack reported on the right)
instead of the one intended ({\small\texttt{bar}}) in a normal execution (stack
on the left).  The figure reports the commands used to run the program in the
two cases. DIFT can prevent this kind of attacks by marking the input of the
program ({\small\texttt{argv}}) as spurious and by enforcing a policy to avoid
using spurious data as pointers. When such violations are detected the
processor raises an exception.
\qed
\end{example}

\begin{figure}[t]
    \centering
    \includegraphics[width=0.34\textwidth]{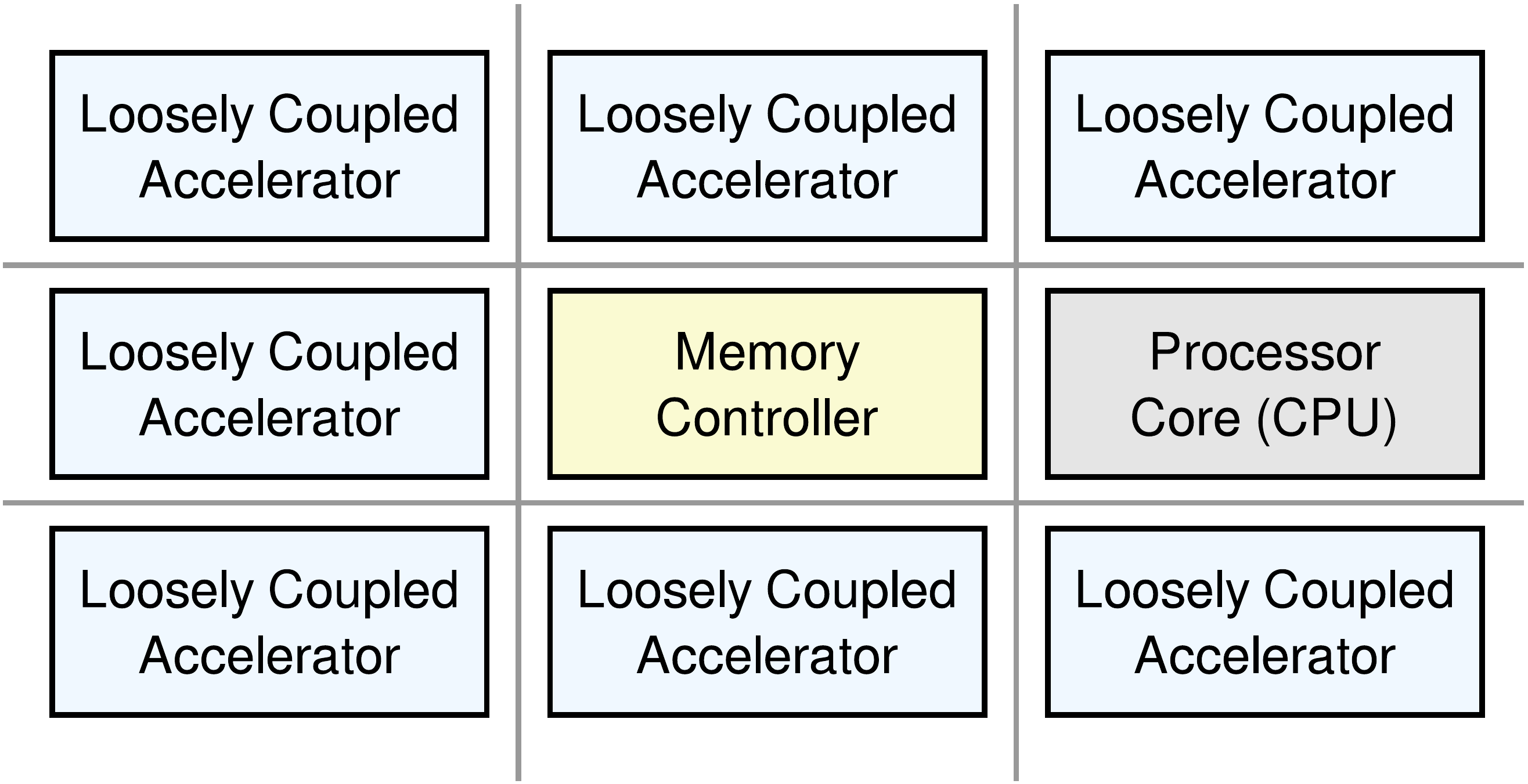}\vspace{-0.15cm}%
    \caption{Architecture of the tile-based systems-on-chip targeted in this work.}
    \label{figure:soc:arch}
\end{figure}

{
Several implementations of DIFT have been proposed in the state of the art, as
reported in Section~\ref{section:related}. Most of these implementations target
only processor cores, rather than entire SoCs. Among them, two schemes can be
used to manage the tags~\cite{porquet2013}. With the \textit{coupled} scheme,
the tag is stored physically with its associated data (same address), i.e., the
memory word is extended to accommodate the tag. Thus, registers, caches and
communication channels are also extended~\cite{crandall2004}. With the
\textit{decoupled} scheme, instead, the tags are stored separately from the
data (different addresses). Typically, the tags are stored in a protected
region in memory~\cite{venka2008}. In our case, as we move to SoCs, we define a
variation of the decoupled scheme where the tags are \textit{interleaved} with
the data. The tags have the same bit width of a memory word. They are inserted
by the operating system and the software programmers remain unaware of their
presence.  With respect to a coupled scheme, an interleaved scheme allows
designers to analyze the effect of changing the \textit{tag offset}, i.e., the
number of words between two consecutive tags in memory. This affects the
security guarantees as well as the performance and cost of the accelerators
(Section~\ref{section:security}). In addition, this scheme does not require a
major modification of the underlying architecture to accommodate the tags.
Thus, in this paper, we focus mainly on such interleaved scheme. To show the
flexibility of our design methodology, however, we present the case study of an
embedded platform that has been extended with DIFT by using a coupled scheme
(Section~\ref{section:results:pulpino}). 
}

%
% Systems-on-Chip and Accelerators
%

%
\subsection{Systems-on-Chip (SoCs) and Accelerators}
\label{section:prelim:arch}

{
\textbf{System-on-Chip Architecture.}
We target a tile-based architecture~\cite{carloni2016} as the one shown in
\figurename~\ref{figure:soc:arch}. Each tile implements a processor core (e.g.,
SPARC V8, RISC-V), a loosely coupled accelerator, or some accessory functionality
such as a memory controller. We assume that the processor core supports DIFT as
described, for example, in~\cite{suh2004}. We aim at extending DIFT to
loosely coupled accelerators by leveraging prior works on processor cores. The
components in our target SoC communicate by means of a network-on-chip or a
bus. The accelerators are managed by the operating system (Linux) trough device drivers.
\fillparagraph
}

\smallskip

\begin{figure}
    \centering
    \includegraphics[width=0.37\textwidth]{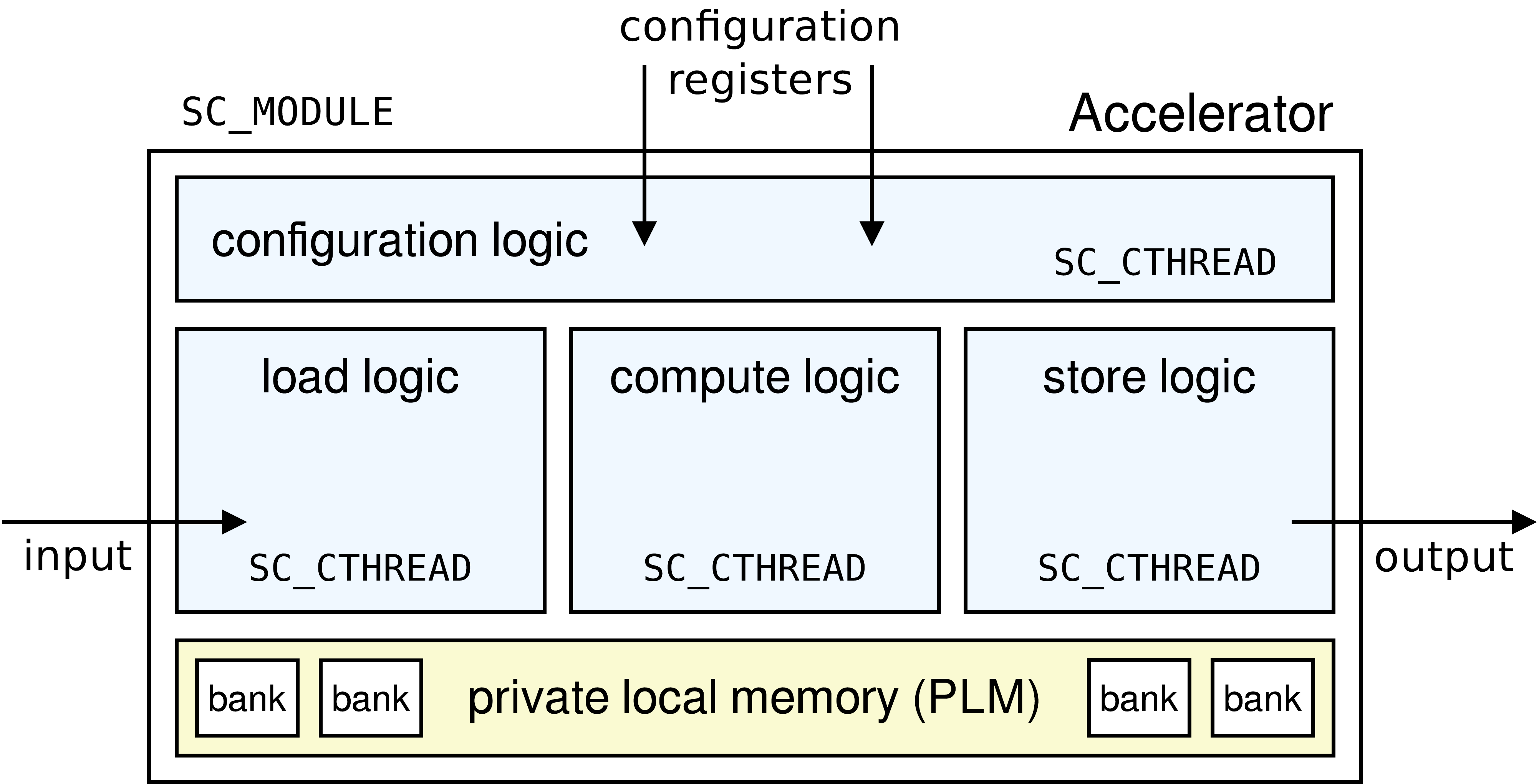}
    \caption{Architecture of the loosely coupled accelerators targeted in this work.}
    \label{figure:accelerator:arch}\vspace{-0.1cm}%
\end{figure}

{
\textbf{Accelerator Architecture.}
This paper focuses on loosely coupled accelerators that have an architecture
similar to the one reported in~\cite{piccolboni2017a}.
We designed our accelerators in SystemC, an IEEE-standard object-oriented
programming language based on {C++}~\cite{systemcref}.
\figurename~\ref{figure:accelerator:arch} shows the architecture, which is
common across all the accelerators we have implemented.  An accelerator is
specified as a SystemC module (i.e., {\small\texttt{SC\_MODULE}}), and the logic
is divided into four components (i.e., {\small\texttt{SC\_CTHREAD}}). The \emph{configuration logic} is used to
setup the accelerator by means of a set of configuration registers.  These
registers are memory mapped, and they are managed by the software application
through the device driver of the accelerator. They define where the input and the output of the
accelerator are in main memory and other parameters that are relevant for the
specific accelerator (e.g., the number of pixels of the images for an
accelerator that processes images).  The \emph{load logic} reads the input data
from main memory by interacting with a DMA controller.  The \emph{store
logic} writes the results of the accelerator back in main memory in a
similar way.  Finally, the \emph{compute logic} implements the specific
computational kernel of the accelerator. The accelerator architecture
includes also a \emph{private local memory} (PLM), or \emph{scratchpad}, which
holds the data during the computation~\cite{pilato2017m, lyons2012}.  PLMs are
usually multi-bank memory architectures that provide multiple read and write
ports to allow accelerators to perform multiple accesses in parallel (in the same clock cycle).  PLMs
occupy a large portion of the accelerator logic, and their size is a key
parameter for design-space exploration. Typically, loosely coupled accelerators
work by dividing the computation into multiple bursts since the workload size 
is much bigger than the capacity of the PLM~\cite{piccolboni2017b}.
Note that several software applications can offload parts of their
computation to the same accelerator at different times. Thus, the PLM is reset
at every invocation of the accelerator to guarantee that a process cannot leak
data from the PLM previously used by another process.
%
%\fillparagraph 
%
}

\smallskip

\begin{figure}
    \centering
    \vspace{0.11cm}%
    \includegraphics[width=0.48\textwidth]{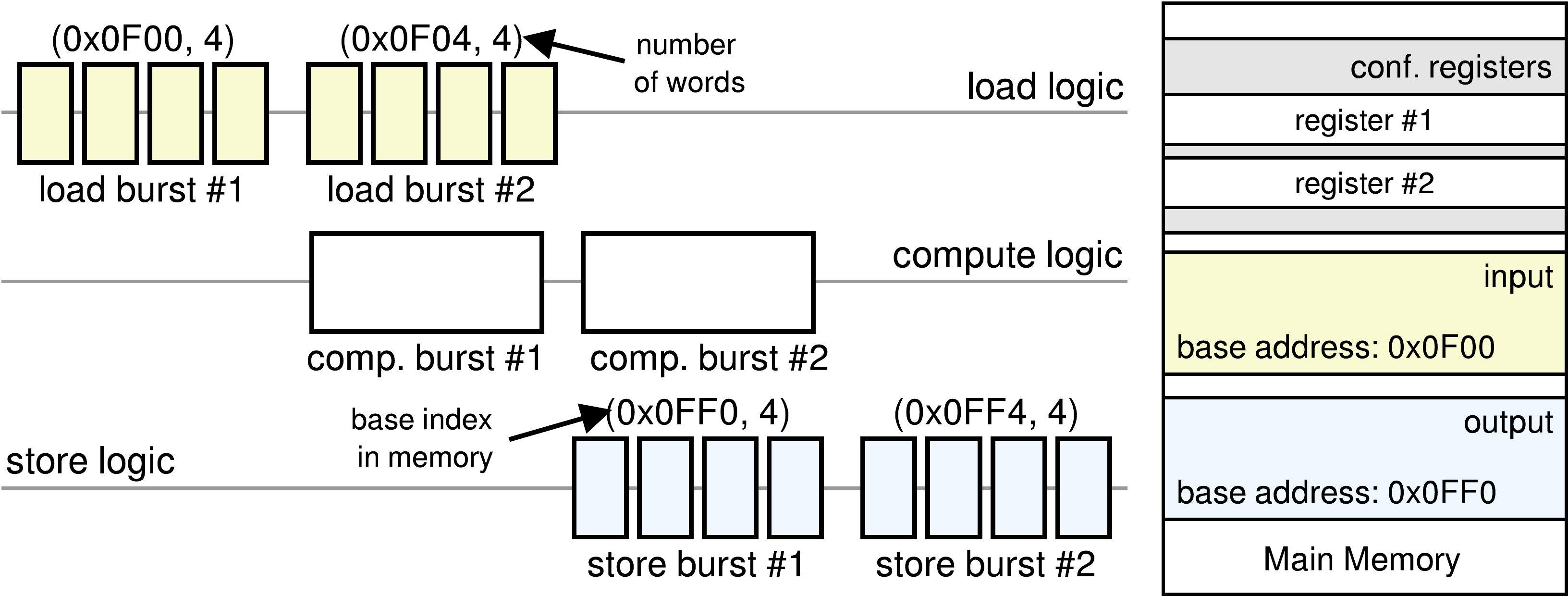}
    \caption{Example of execution of a loosely coupled accelerator.}
    \label{figure:accelerator:exec}\vspace{-0.04cm}%
\end{figure}

{
\textbf{Accelerator Execution.}
To offload the computation to an accelerator, its device driver is called.  The
software application prepares the input data for the accelerator in main memory
and uses the Linux system call {\small\texttt{ioctl()}} to invoke the device
driver.  The device driver writes the memory-mapped registers and runs the
accelerator.  The accelerator raises an interrupt after completing its
execution so that the processor can resume the execution of the software
application.
An example of accelerator execution is shown in
\figurename~\ref{figure:accelerator:exec}, which also reports the layout in
main memory of the accelerator data. The accelerator first loads a subset of
the input data by operating with bursts of a fixed length, called \emph{burst
size}. The accelerator loads the data autonomously into its PLM via DMA,
without any intervention of the processor core. Each burst is defined by the
index in memory from which the data must be read (or must be written to) and
the length of the burst in terms of memory words (in
\figurename~\ref{figure:accelerator:exec}, these values are indicated as pairs
above each burst). The burst size is limited by the amount of data that can be
stored in the PLM, and it has been shown to be important for design-space
exploration~\cite{piccolboni2017b}. Then, the accelerator computes the results
for the given load burst, and it writes them into main memory with a store
burst. These three phases can be pipelined by increasing the PLM size. For a
given accelerator, it is not necessarily the case that each load burst is
followed by one store burst.  Some accelerators need multiple load bursts to
produce one store burst. For instance, in the case of matrix multiplication,
the accelerator needs to load one row from the first input matrix and all the
columns from the second input matrix in order to calculate a single row of the
output matrix (assuming that the burst size corresponds to the size of a row).
%
%\fillparagraph
%
}

%
% Assumptions and Attack Model 
%

%
\subsection{Assumptions and Attack Model}
\label{section:prelim:attmodel}

{
We make the following assumptions regarding the hardware architecture, the
software environment and the attacker's capabilities\footnote{{We assume that a
hardware implementation of DIFT is available for the processor and the
communication infrastructure. A equally valid alternative would be having a
hybrid approach where the accelerators are protected in hardware while the
software applications are protected by a software-based DIFT approach within
the operating system (see Section~\ref{section:related} for related work).}}.
We assume to have a processor extended with DIFT and that the hardware
(including the accelerators) is trusted, i.e., no hardware Trojans. We assume
that the communication infrastructure that connects the hardware components
(\figurename~\ref{figure:soc:arch}) supports the tags. We aim at extending the
security guarantees provided by DIFT to accelerators.  The architecture may
include some common hardware defenses, e.g., non-executable memory. 
In this paper we address software-based attacks, e.g. buffer-overflow attacks,
return-to-libc attacks, etc. We target user-space applications that offload
parts of their computation to accelerators. The applications are executed
either in the context of the Linux operating system
(Section~\ref{section:results}) or in bare metal
(Section~\ref{section:results:pulpino}). In the first case, we assume that the
device drivers run in kernel space, or that they are trusted if they run in
user space. We assume that the applications have one or more vulnerabilities,
e.g., buffer overflows, format string bugs, etc. The attacker exploits these
vulnerabilities through common I/O interfaces, with the goal of affecting the
integrity and/or confidentiality of the hardware-accelerated software
applications. 
\fillparagraph
}

%
% Motivational example 
%

%
\section{Need of a Holistic DIFT Implementation}
\label{section:example}

{
Heterogeneous SoCs consist of multiple processor cores and accelerators. To
guarantee the security of such systems with DIFT, we need to implement a
\textit{holistic approach}: DIFT must be supported in both processors and
accelerators. This ensures that (1) the tags are propagated from the processor
cores to the accelerators and vice versa, and (2) the policies are enforced
(i.e., the tags are checked) in both processors and accelerators.
\fillparagraph
}

\begin{figure}
    \centering
    \vspace{0.05cm}%
    \includegraphics[width=0.47\textwidth]{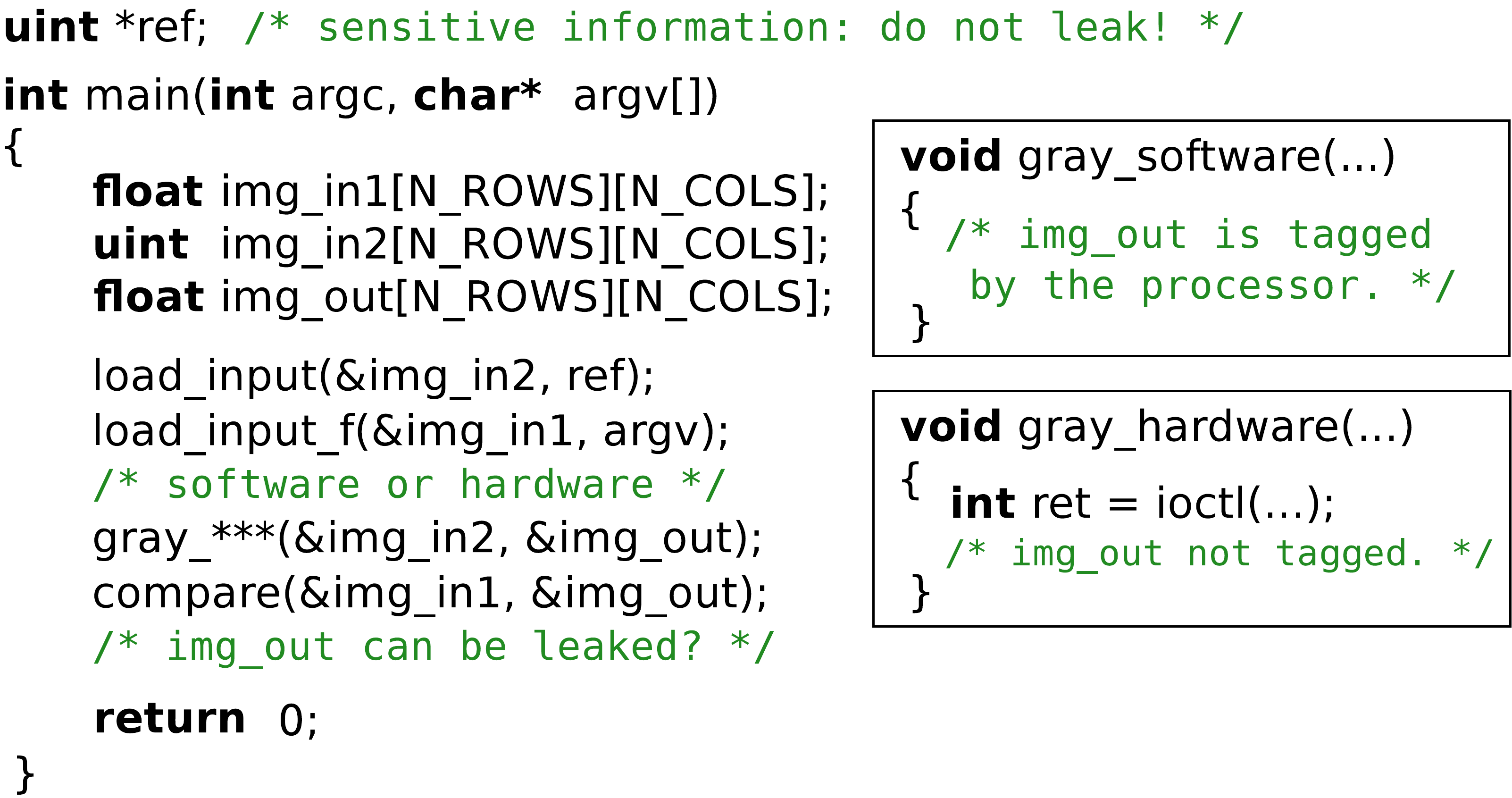}\vspace{-0.04cm}%
    \caption{Example of leakage of information if {\scriptsize \texttt{gray}} is executed in hardware.}
    \label{figure:attack}\vspace{-0.09cm}%
\end{figure}

\begin{example}
Consider the code reported in \figurename~\ref{figure:attack} that can be used,
for example, in a video-surveillance \mbox{system.  Suppose} that
{\small\texttt{ref}} contains a face image that is compared with the image
passed through {\small\texttt{argv}}. We want to enforce a DIFT policy that
ensures that {\small\texttt{ref}} cannot be leaked for any reason. Before the
comparison, {\small\texttt{ref}} is converted to the same format of the input
image, e.g., from RGB to grayscale. The function for the conversion is
initially implemented in software ({\small\texttt{gray\_software}}). The
processor, which supports DIFT, guarantees that when this image is manipulated
it is properly tagged. In other words, the tags are propagated and no leaks are
possible.  Suppose that the conversion is now implemented with an equivalent
accelerator to improve performance ({\small\texttt{gray\_hardware}}). If the
accelerator is not extended with DIFT, {\small\texttt{ref}} is vulnerable to
leaks. \qed
\fillparagraph
\end{example}

{
%
%Extending the support of DIFT to accelerators is thus critical to avoid such
%security exploits. We explain how we can protect the accelerators execution
%with DIFT in the following section.
%
}

%
% The DIFT Shell for Accelerators
%

%
\section{DIFT Shell for Accelerators}
\label{section:design}
%

%
%We designed the DIFT shell to be \textit{double decoupled} with respect to the
%accelerator, and to be \textit{flexible}. In this section, we first discuss
%these implementation choices. Then, we describe the shell architecture and how
%it encapsulates the accelerator.
%

%
We designed the DIFT shell to be \textit{double decoupled} with respect to the
accelerator, and to be \textit{flexible}. In this section, we first discuss the
shell architecture and how it encapsulates the accelerator. Then, we discuss
such design choices.
%
%
% The Implementation of the DIFT Shell
%

\vspace{-0.1cm}
\subsection{Implementation of the DIFT Shell}
\label{section:design:impl}

\begin{figure}[t]
    \centering
    \includegraphics[width=0.45\textwidth]{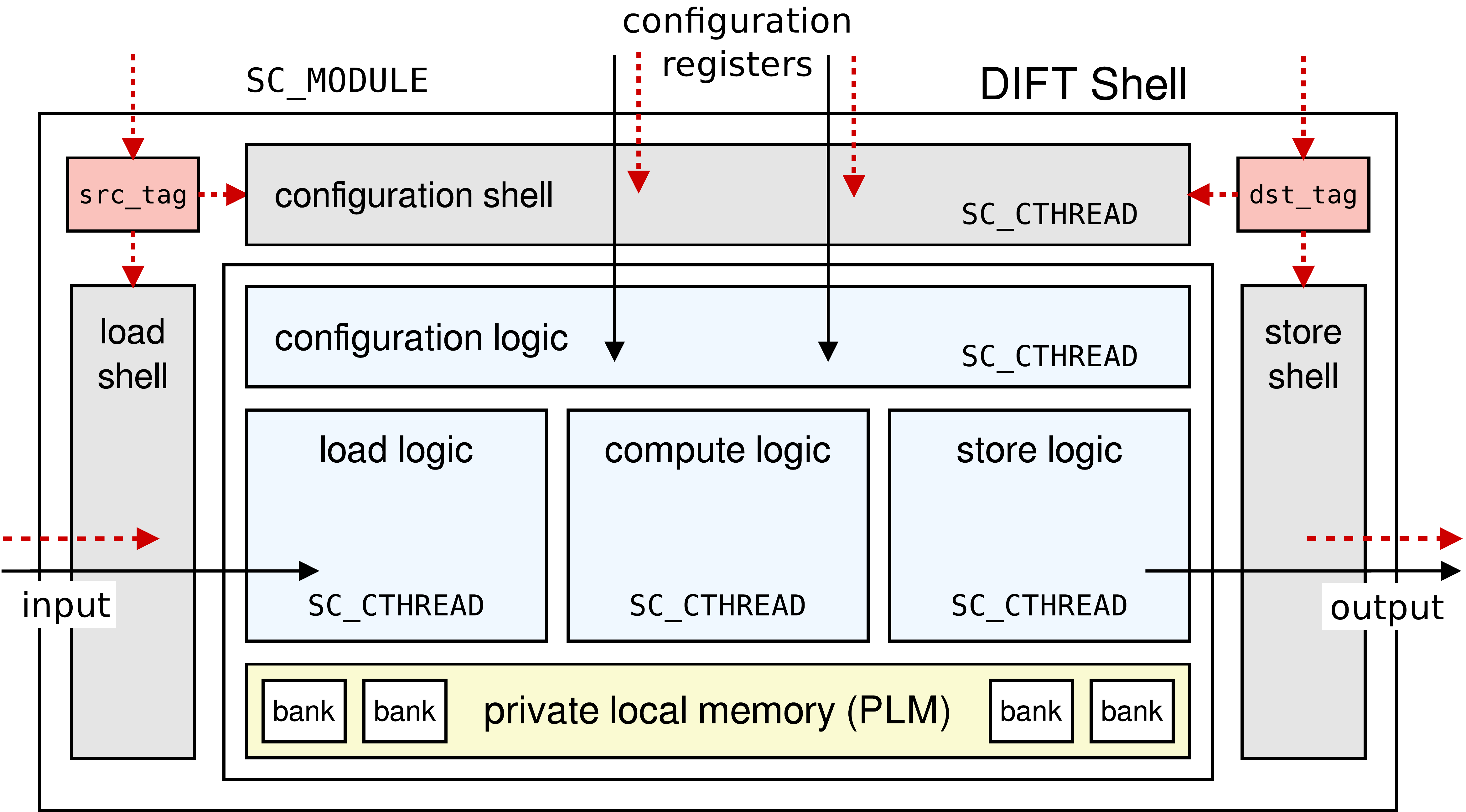}
    \caption{Architecture of the DIFT shell.}\vspace{-0.17cm}%
    \label{figure:shell:arch}
\end{figure}

{
\textbf{Shell Architecture.}
We designed the DIFT shell in {SystemC} with an architecture similar to the
accelerators. \figurename~\ref{figure:shell:arch} shows how the shell
encapsulates the accelerator and distinguishes the data flows (black solid
arrows) from the tag flows (red dashed arrows).  The logic is divided into
three main components. The {configuration logic} (\textit{configuration shell}
in \figurename~\ref{figure:shell:arch} to distinguish it from the configuration
logic of the accelerator of \figurename~\ref{figure:accelerator:arch}) sets up
the shell through a set of configuration registers. The shell has {\small $2
\times N + 2$} memory-mapped registers, where {\small $N$} is the number of
registers {of the} accelerator.  Each register of the accelerator is tagged to
ensure that it cannot be easily compromised by an attacker since it can contain
sensitive information such as the addresses in main memory where the inputs and
outputs of the accelerator reside. Note that the registers containing the tags
are not visible to the software applications and they are managed by the device
driver. We have also two additional configuration registers.  The register
{\small\texttt{src\_tag}} is used to specify the value of the tags interleaved
in the input of the accelerator and in the configuration registers.  The
register {\small\texttt{dst\_tag}} has the value of the tags to be interleaved
in the output of the accelerator.  The values of these two registers are not
visible to the software applications.  They are managed by the device driver of
the accelerator.  In particular, these values are passed from the processor
when the device driver is called. In this way the tags can be propagated
similarly to the case in which the accelerator functionality is executed in
software. This is a form of coarse-grain DIFT, where the output tags of the
accelerators are determined solely on the basis of the input tags\footnote{In
Section~\ref{section:discussions} we compare coarse-grain and fine-grain DIFT
approaches.}.  There are other two components in the shell of
\figurename~\ref{figure:shell:arch}.  The {load logic} (\textit{load shell})
receives the read requests of the accelerator and it modifies them to consider
the tags, i.e., the shell modifies the base address in memory and the length of
the requests to include the tags if necessary. Then, it passes the values to
the accelerator while checking that the tag values interleaved with input data in
main memory match the value specified in {\small\texttt{src\_tag}}.  In case of
mismatch, it immediately stops the execution of the accelerator.  The {store
logic} (\textit{store shell}) intercepts the write requests of the accelerator
in a similar way and writes the results of the accelerator by interleaving the
tags with the value in {\small\texttt{dst\_tag}}. 
\fillparagraph
}

\smallskip

{
\textbf{Tag Interleaving.}
We use an interleaved scheme to handle the tags
(Section~\ref{section:prelim:dift}): the tags are interleaved in memory with a
fixed tag offset. However, interleaving the tags uniformly in memory by
starting always from the same location is not secure. An attacker could infer
the locations of the tags, for instance by executing the accelerator
repeatedly.  The attacker could then replace the input of the accelerator with
malicious data by skipping the tag locations. Therefore, in our implementation
of the DIFT shell, we keep a fixed distance (in words) between two consecutive
tags in memory, but we randomize the location of the first tag interleaved in
the input data at every execution of the accelerator to make it not
predictable.  The DIFT shell needs to know the offset of the first tag embedded
in the input in order to check the tags and pass only the data values to the
accelerator. To do that, we add another configuration register to the shell of
\figurename~\ref{figure:shell:arch}. The location of the first tag is generated
with a pseudo-random number generator.  We chose to use a single tag value for
the input ({\small\texttt{src\_tag}}) and a single tag value for the output
({\small\texttt{dst\_tag}}) of the accelerator.  However, (1) the pattern with
which the tags are interleaved in the input and output of the accelerator and
(2) the number of different values for the tags can be customized to offer
stronger security guarantees without requiring any modification to the
accelerator implementation.  Alternatively, it is possible to use
data-dependent tags instead of a randomized approach, i.e., the values of
{\small\texttt{src\_tag}} and {\small\texttt{dst\_tag}} can be calculated by
applying a crypto hash function to the inputs and outputs of a specific
accelerator execution. 
}

\smallskip

\begin{figure}
    \centering
    \includegraphics[width=0.46\textwidth]{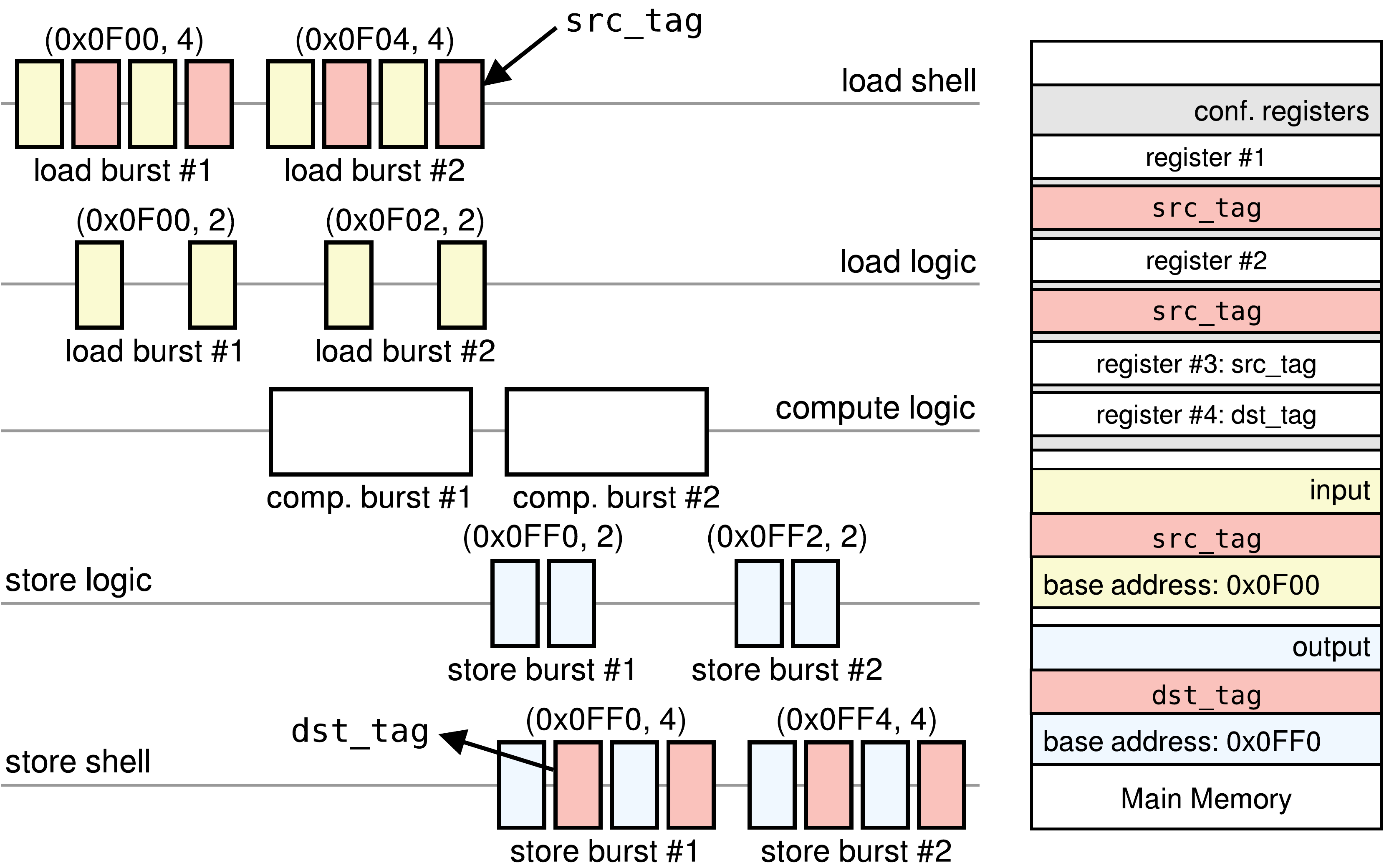}
    \caption{Example of execution of the DIFT shell.}\vspace{-0.15cm}%
    \label{figure:shell:exec}
\end{figure}

{ \textbf{Shell Execution.}
\figurename~\ref{figure:shell:exec} shows an example of execution of the DIFT
shell encapsulating an accelerator.  We consider the case where the tag offset
is equal to one, i.e., the size of the input and output of the accelerator
is doubled to add to each value a tag in the next memory location.  This corresponds
to the case where we have the maximum number of tags in memory. The load
requests of the accelerator consist of two memory words. The shell modifies
such requests by doubling the amount of data to include the tags. While doing
so, the shell verifies that the tags from the main memory contain the value
specified in {\small\texttt{src\_tag}}.  Similarly, the store requests from the
accelerator (two memory words) are modified to interleave the tags with
the value in {\small\texttt{dst\_tag}}, thus marking the outputs. 
}

%
% The Adaptability of the DIFT Shell
%

%
\subsection{Adaptability of the DIFT Shell}

{
\textbf{Double Decoupling.}
Inspired by the principles of latency-insensitive design~\cite{carloni2015}, we
designed the DIFT shell to be double decoupled from the accelerator
implementation, i.e., the design of the accelerator is independent from the
design of the shell, and the design of the shell is independent from the design
of the accelerator.  Besides the I/O interface, the shell needs to know only
the number of configuration registers of the accelerator, which is usually
decided at design time. This allows designers to rapidly integrate DIFT in
their accelerators.  Designers can also easily extend third-party intellectual
property (IP) cores with DIFT in their SoCs, simplifying the implementation of
a holistic DIFT approach. For example, the generation of the shell and the
connection with the accelerator, in our implementation, is done automatically.
This design choice makes the design of the shell be independent from the
accelerators design as well, which guarantees the reusability of the shell.
Note that the tags are not propagated into the logic of the accelerators, which
remain \textit{completely unaware} of the tags. This guarantees a minimal area
overhead, but it could limit the set of policies a designer may want to
support, as discussed in Section~\ref{section:discussions}.
}

\smallskip

{
\textbf{Flexibility.}
The DIFT shell is flexible because the interface to communicate with the
network-on-chip or bus (Section~\ref{section:prelim:arch}) is decoupled from
the internal logic.  In addition, the shell and the accelerator expose the same
interface, allowing designers to easily replace an accelerator with its
encapsulated version. Note also that the shell can be easily customized to
the needs of the specific accelerator, for example to (1) improve performance
or (2) strengthen security. In Section~\ref{section:design:impl} we presented
an implementation of the shell that uses an interleaved scheme for the
tags. The shell can be adapted to work with different schemes as well.  We show
an example of this customization in
Section~\ref{section:results:pulpino}.
\fillparagraph
}

%
% A Security Metric 
%

\section{A Security Metric for Accelerators}
\label{section:security}

{
In this section we define a security metric for accelerators to quantitatively
evaluate the security guarantees provided by the DIFT shell. This metric is a
valuable parameter for a multi-objective design-space exploration of
accelerators, where not only performance and cost but also security is a
critical aspect.
}

%
% Metric Definition 
%

\subsection{Information Leakage: Metric Definition}

%%%% Definition

\begin{definition}
\label{def:info_leakage}
The information leakage is the amount of data that can be produced as
output by an accelerator before its shell realizes that the input has
been corrupted by an attacker.
\end{definition}

%%%% Example

\begin{figure}
    \centering
    \includegraphics[width=0.45\textwidth]{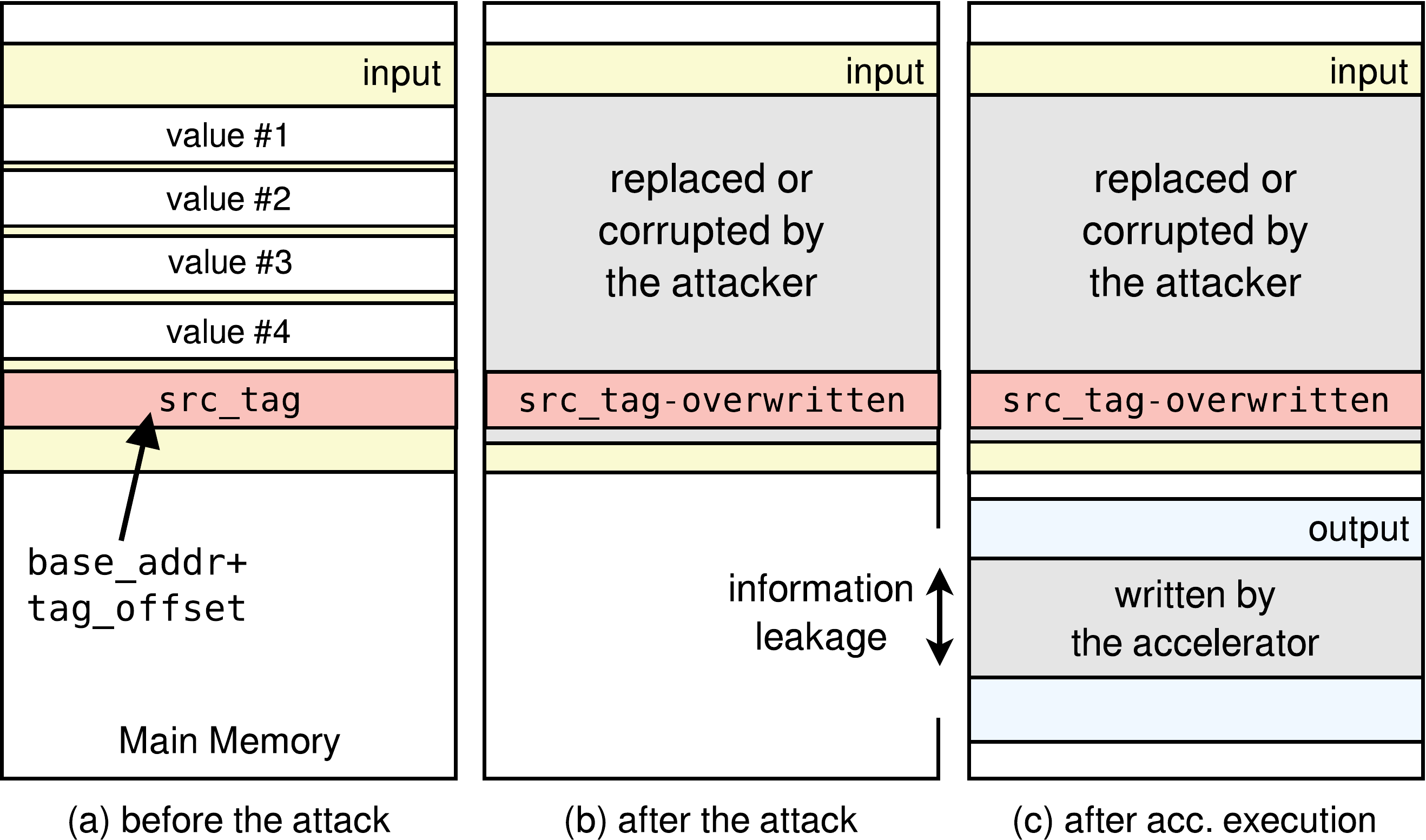}\vspace{-0.1cm}%
    \caption{The metric of information leakage for loosely coupled accelerators.}
    \label{figure:leakage}
\end{figure}

\begin{example}
\label{example:leakage} 
To calculate the information leakage for a given accelerator execution, we
consider the worst-case scenario: the first tag is inserted in the farthest
location in main memory, according to the value of the tag offset, from the
beginning of the input data of the accelerator.  In other words, the tag is at
the memory location with address {\small$\texttt{base\_addr} +
\texttt{tag\_offset}$}, where {\small\texttt{base\_addr}} is the first address
in main memory where the input of the accelerator is stored. This scenario is
depicted in \figurename~\ref{figure:leakage}~(a). An attacker could try to
corrupt the input data of the accelerator in memory. However, in doing so, the
attacker would inadvertently overwrite the first tag as well
(\figurename~\ref{figure:leakage}~(b)). In fact, the attacker cannot easily
determine the exact distance between two consecutive tags in memory and the
initial offset of the tags, thanks to the randomized approach we adopted for
the DIFT shell (Section~\ref{section:design}). Thus, the information leakage is
the percentage of output values (produced by the accelerator before the shell
realizes that it has been compromised) with respect to the total amount of
values the accelerator would generate if it was not compromised. This
corresponds to the amount of output produced by the accelerator before the
shell stops its execution (\figurename~\ref{figure:leakage}~(c)). The shell
realizes that it has been compromised when it reads the first tag, which has
been overwritten by the attacker\footnote{In this example, without loss of
generality, we are assuming that the accelerator starts to read the input from
the first memory location ({\scriptsize\texttt{base\_addr}}).}. This
calculation represents an upper bound to the information leakage. Note that the
same reasoning can be applied when the attacker tries to corrupt the input of
the accelerator not by starting from {\small\texttt{base\_addr}}.
\qed
\end{example}

{
The generic security concept of ``information leakage'' has been adapted to our context with
Definition~\ref{def:info_leakage}. Note that this definition applies to the decoupled
scheme that we discussed in Section~\ref{section:prelim:dift}, where the tags
are interleaved in memory with the data. This definition does not apply to the
coupled scheme since each data is stored with its tag. For such a scheme we can
adopt the concept of \textit{security proportionality}~\cite{porquet2013}.  Next,
we describe how information leakage is influenced \mbox{by different factors.}
}

%
% Metric Analysis
%

\subsection{Information Leakage: Metric Analysis}

{
We perform a {quantitative information-flow analysis}~\cite{kopf2013} to
measure the information leakage caused by the accelerators when protected 
with the {DIFT} shell. In our analysis, we found that information 
leakage depends on the following factors.
}

\smallskip

% -----------------------------------------------------------------------------

{
\noindent\textbf{(1) Tag Density.}
The more tags we interleave in the input data of the accelerator, the higher is
the likelihood that the shell hits a tag that has been corrupted before
producing an output.
}

\smallskip

{
The information leakage is expected to decrease as the tag density increases.
The tag density is thus a key parameter for a multi-objective design-space
exploration that considers cost, performance, and security of accelerators. In
fact, by increasing the number of tags in memory we guarantee potentially a
lower information leakage because it is more likely that the attacker replaces
a tag (as shown in \figurename~\ref{figure:leakage}). This, however,
negatively affects the performance (more input/output to process) and cost
(overhead for the tags in memory) of executing the accelerator.
}

\smallskip

% -----------------------------------------------------------------------------

{
\noindent\textbf{(2) Algorithm.}
The algorithm implemented by the accelerator defines the amount of inputs
needed to calculate an output. This affects the memory access pattern,
and thus the security.
}

\begin{example}
Consider two algorithms: (1) image conversion from RGB to grayscale values, and
(2) matrix multiplication. For (1) to calculate one grayscale value we need
only the RGB value in the corresponding position in the input image. For (2),
instead, we need to load a row and a column of the input matrices to calculate
a single value of the output matrix.
\qed
\fillparagraph
\end{example}

{
The information leakage is expected to be higher for those accelerators that
require fewer input values to calculate the corresponding output value. In fact,
accelerators usually work in bursts. If an accelerator needs fewer input to
calculate an output, a lower number of load bursts is required to produce a
corresponding store burst. Thus, it is more likely that the accelerator 
produces outputs before the shell finds an invalid tag.
\fillparagraph
}

\smallskip

% -----------------------------------------------------------------------------

{
\noindent\textbf{(3) Implementation.}
The specific way in which the accelerator implements the 
algorithm can affect the information leakage. 
}

\begin{example}
\label{example:burstsize}
Consider an accelerator performing the conversion from RGB to grayscale. If it
operates in bursts of 16 pixels, each load burst of 16 pixels produces a store
burst of 16 pixels (for an efficient use of the PLM).  Similarly, if the
accelerator uses bursts of 1024 pixels, each load of 1024 pixels produces a
store burst of 1024 pixels. Assume there is one tag every 1024 pixels in the
input image in main memory: in the first case, the accelerator leaks data
before encountering a tag that has been compromised (in the worst-case
scenario), while, in the second case, the accelerator does not cause leakage.
\qed
\end{example}

{
The information leakage is expected to decrease as the burst size increases.
In fact, the larger are the bursts, the higher is the probability of finding a
tag. The burst size affects performance and cost as well. Larger bursts require
larger PLMs for storing data during the computation (higher cost), but they
improve the efficiency of data transfers because few large bursts via DMA are
generally much more efficient than many small bursts~\cite{piccolboni2017b}.
}

\smallskip

% -----------------------------------------------------------------------------

{
\noindent\textbf{(4) Workload Size.}
The workload size determines the total amount of data the accelerator needs to
process. This affects the memory access pattern, and thus the information leakage. 
\fillparagraph
}

\smallskip

{
If the accelerator works on relatively small inputs, the input set can
be entirely stored in the PLM. The accelerator does not need to work in bursts
and it cannot cause information leakage (if there is at least one tag in the
input). If the accelerator works on relatively large inputs (the common case
for loosely coupled accelerators), then it is necessary to work in bursts, and
the leakage is affected by the burst size as described in 
Example~\ref{example:burstsize}.
\fillparagraph
}

%
% RESULTS: leakage
%

\begin{figure*}[!tbp]
%
% MEAN
%
  \centering
  \vspace{-0.1cm}%
  \begin{minipage}[t]{0.24\textwidth}
    \includegraphics[width=\textwidth, trim = {0.95cm 1.1cm 0.35cm 0}]{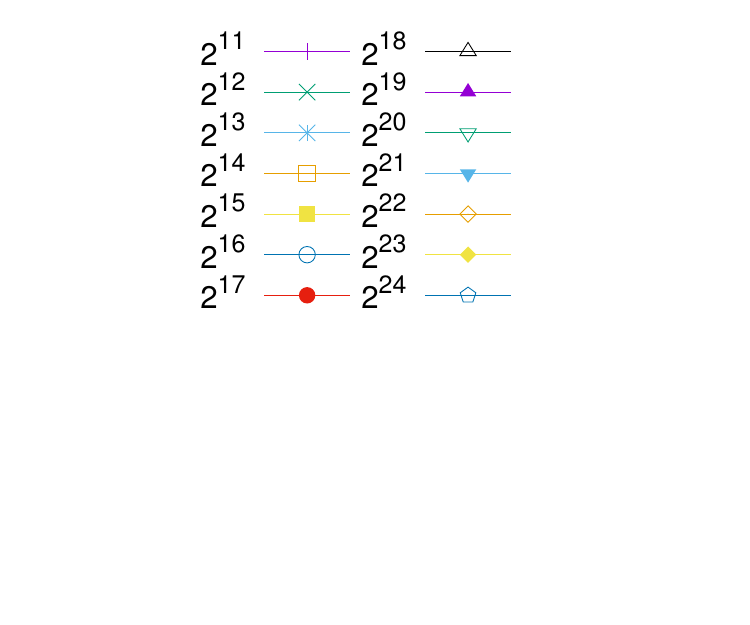}
  \end{minipage}
  \hspace{-1.29cm}%
  \begin{minipage}[t]{0.26\textwidth}
    \includegraphics[width=\textwidth, trim = {0.25cm 0.3cm 0.25cm 0}]{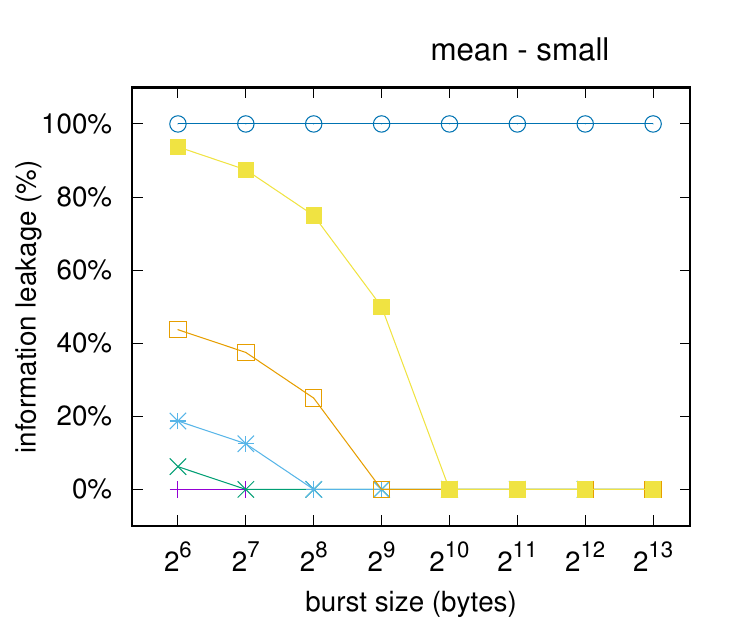}
  \end{minipage}
  \hspace{-0.3cm}%
  \begin{minipage}[t]{0.26\textwidth}
    \includegraphics[width=\textwidth, trim = {0.25cm 0.3cm 0.25cm 0}]{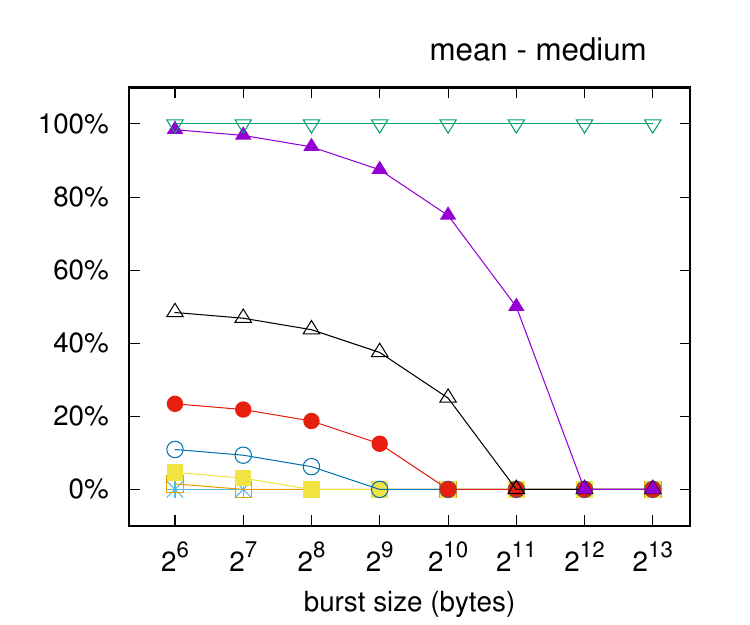}
  \end{minipage}
  \hspace{-0.3cm}%
  \begin{minipage}[t]{0.26\textwidth}
    \includegraphics[width=\textwidth, trim = {0.25cm 0.3cm 0.25cm 0}]{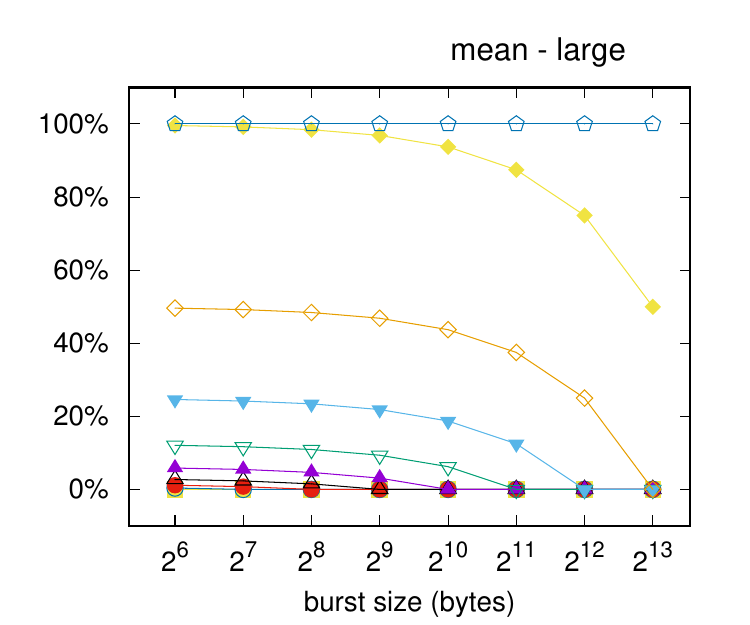}
  \end{minipage}

%
% GRAY
%
  \centering
  \begin{minipage}[t]{0.2\textwidth}
    \includegraphics[width=\textwidth, trim = {0.45cm 1.2cm 0.75cm 0}]{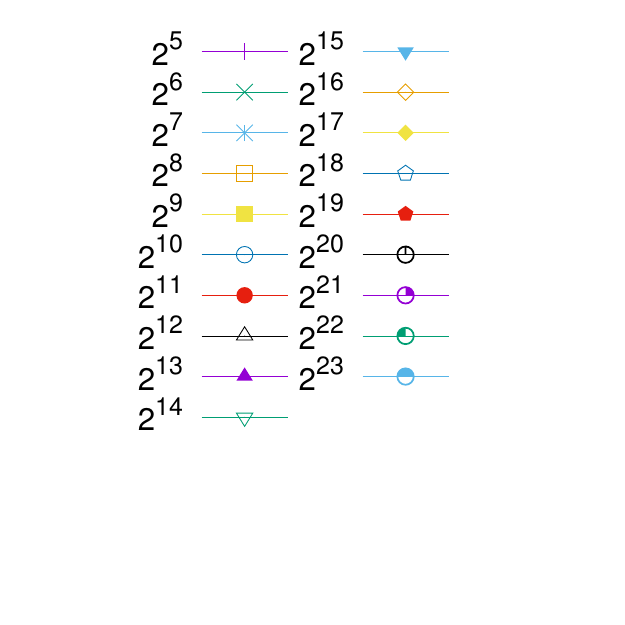}
  \end{minipage}
  \hspace{-0.6cm}%
  \begin{minipage}[t]{0.26\textwidth}
    \includegraphics[width=\textwidth, trim = {0.25cm 0.3cm 0.25cm 0}]{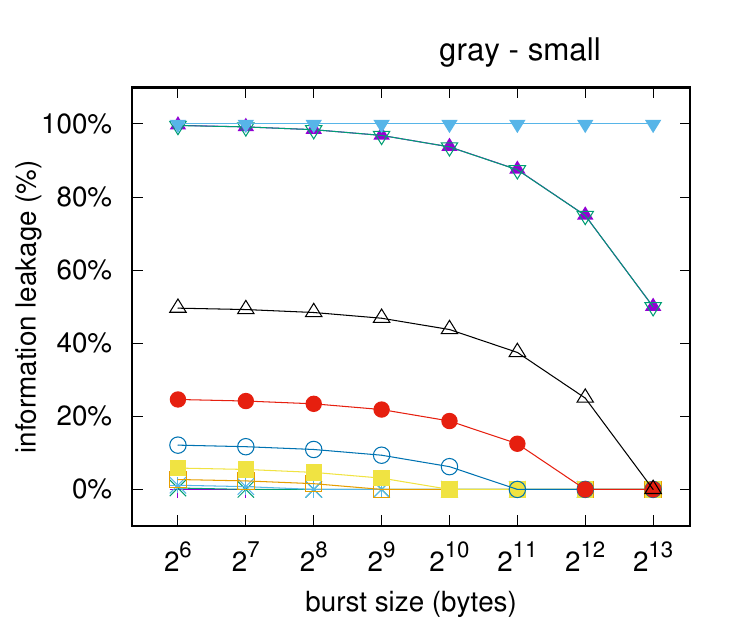}
  \end{minipage}
  \hspace{-0.3cm}%
  \begin{minipage}[t]{0.26\textwidth}
    \includegraphics[width=\textwidth, trim = {0.25cm 0.3cm 0.25cm 0}]{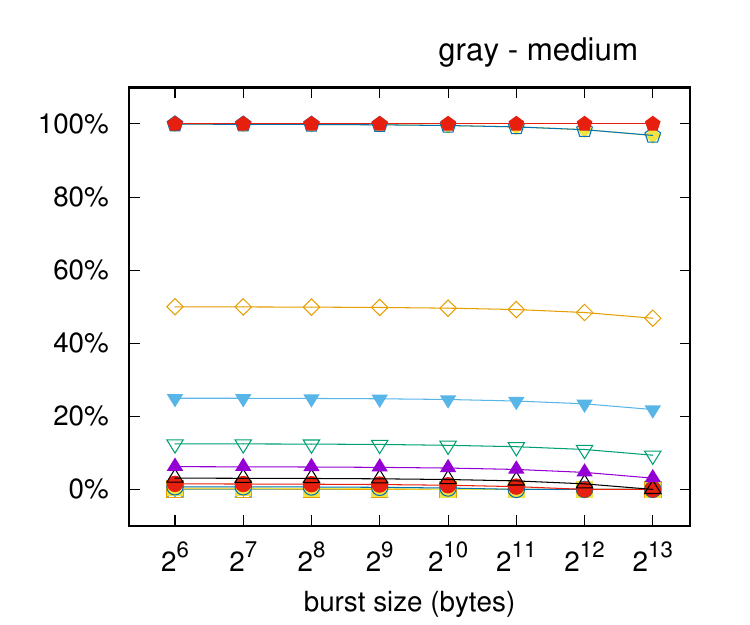}
  \end{minipage}
  \hspace{-0.3cm}%
  \begin{minipage}[t]{0.26\textwidth}
    \includegraphics[width=\textwidth, trim = {0.25cm 0.3cm 0.25cm 0}]{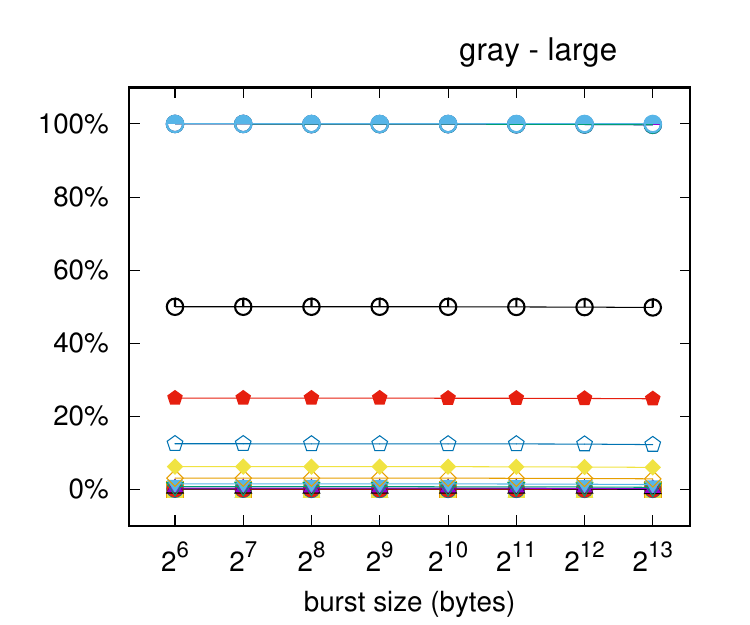}
  \end{minipage}

%
% MULTS
% 
  \centering
  \begin{minipage}[t]{0.24\textwidth}
    \includegraphics[width=\textwidth, trim = {0.95cm 0.9cm 0.35cm 0}]{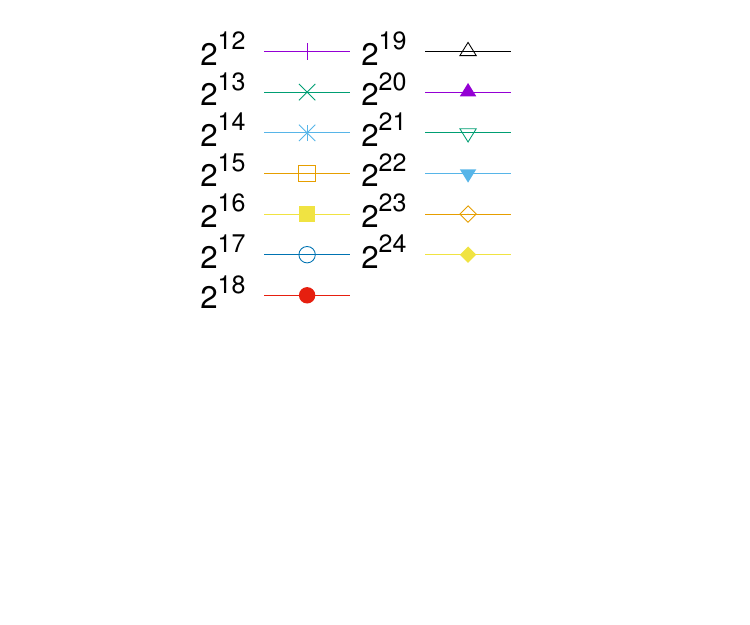}
  \end{minipage}
  \hspace{-1.29cm}%
  \begin{minipage}[t]{0.26\textwidth}
    \includegraphics[width=\textwidth, trim = {0.25cm 0 0.25cm 0}]{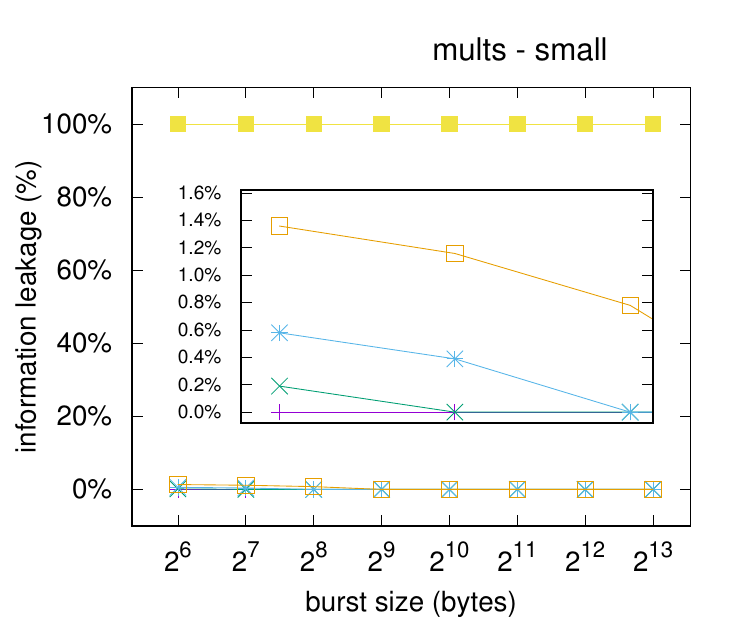}
  \end{minipage}
  \hspace{-0.3cm}%
  \begin{minipage}[t]{0.26\textwidth}
    \includegraphics[width=\textwidth, trim = {0.25cm 0 0.25cm 0}]{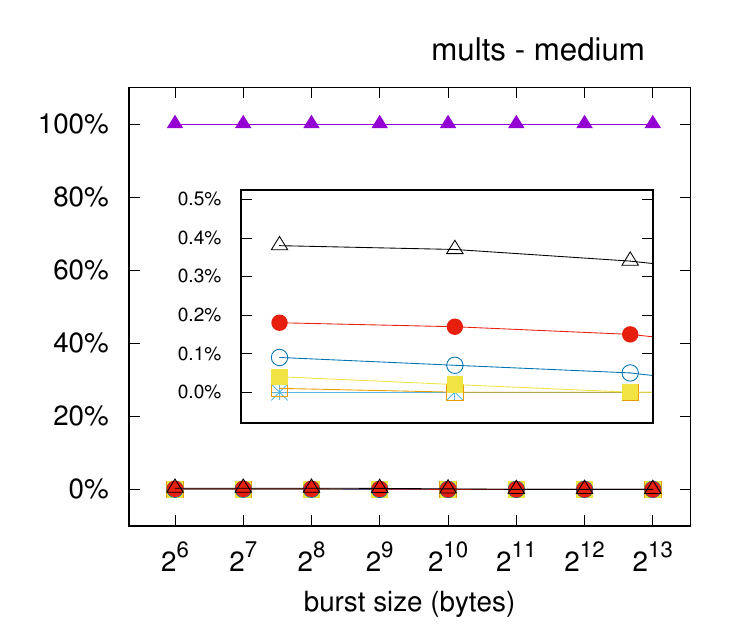}
  \end{minipage}
  \hspace{-0.3cm}%
  \begin{minipage}[t]{0.26\textwidth}
    \includegraphics[width=\textwidth, trim = {0.25cm 0 0.25cm 0}]{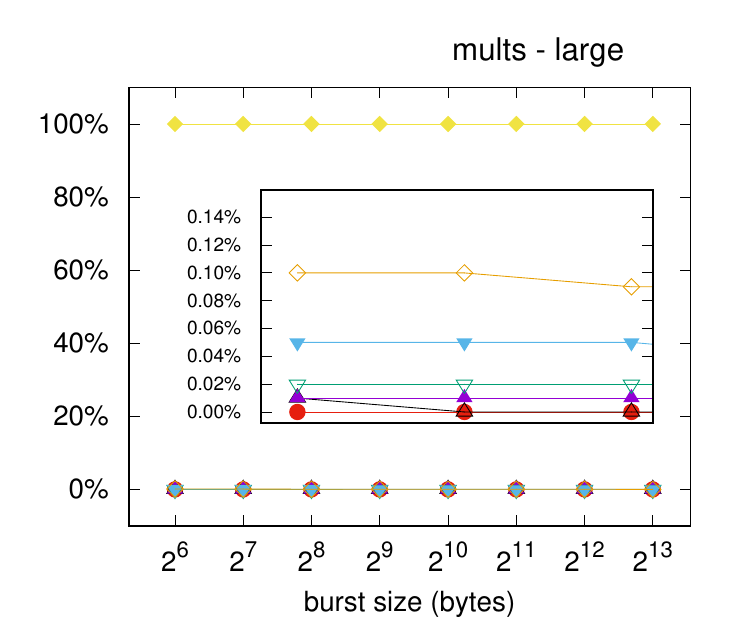}
  \end{minipage}
%
% CAPTION
%
  \caption{Measurements of information leakage for the three accelerators and 
   workloads reported in~\tablename~\ref{table:workloads}. The legend indicates
   the values of the tag offset.}\label{figure:leakageres}
  \vspace{-0.3cm}%
\end{figure*}

%
% RESULTS: space overhead 
%

\begin{figure*}[!tbp]
%
% MEAN
%
  \centering
  \vspace{-0.1cm}
  \begin{minipage}[t]{0.24\textwidth}
    \includegraphics[width=\textwidth, trim = {0.95cm 1.1cm 0.35cm 0}]{plots/info_leakage/mean/mean_legend.pdf}
  \end{minipage}
  \hspace{-1.29cm}%
  \begin{minipage}[t]{0.26\textwidth}
    \includegraphics[width=\textwidth, trim = {0.25cm 0.3cm 0.25cm 0}]{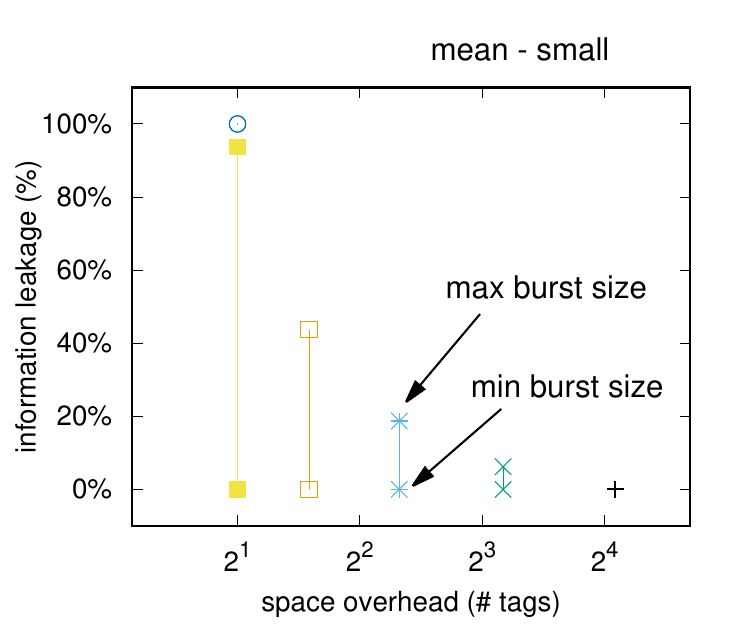}
  \end{minipage}
  \hspace{-0.3cm}%
  \begin{minipage}[t]{0.26\textwidth}
    \includegraphics[width=\textwidth, trim = {0.25cm 0.3cm 0.25cm 0}]{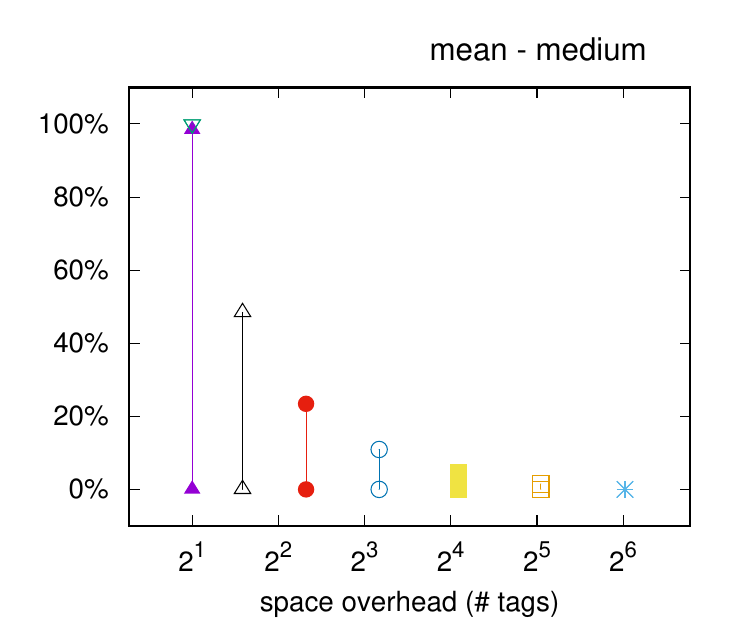}
  \end{minipage}
  \hspace{-0.3cm}%
  \begin{minipage}[t]{0.26\textwidth}
    \includegraphics[width=\textwidth, trim = {0.25cm 0.3cm 0.25cm 0}]{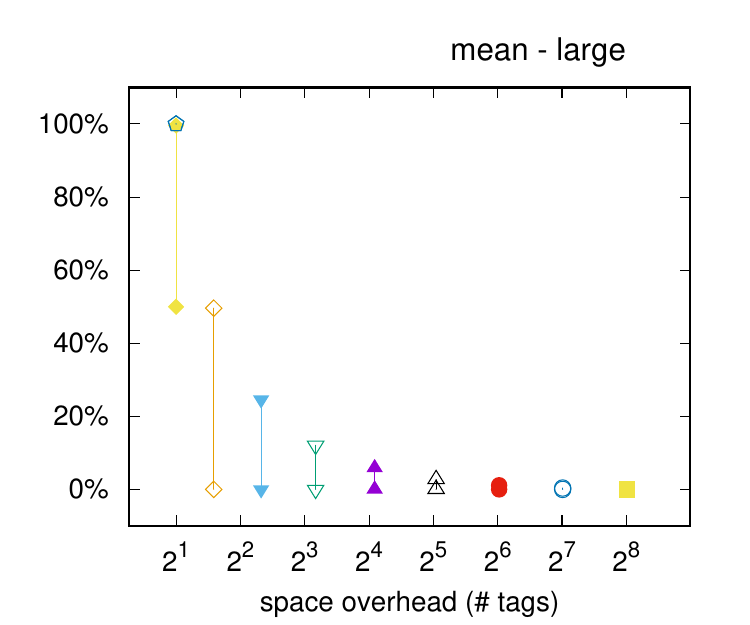}
  \end{minipage}

%
% GRAY
%
  \centering
  \begin{minipage}[t]{0.2\textwidth}
    \includegraphics[width=\textwidth, trim = {0.45cm 1.2cm 0.75cm 0}]{plots/info_leakage/gray/gray_legend.pdf}
  \end{minipage}
  \hspace{-0.6cm}%
  \begin{minipage}[t]{0.26\textwidth}
    \includegraphics[width=\textwidth, trim = {0.25cm 0.3cm 0.25cm 0}]{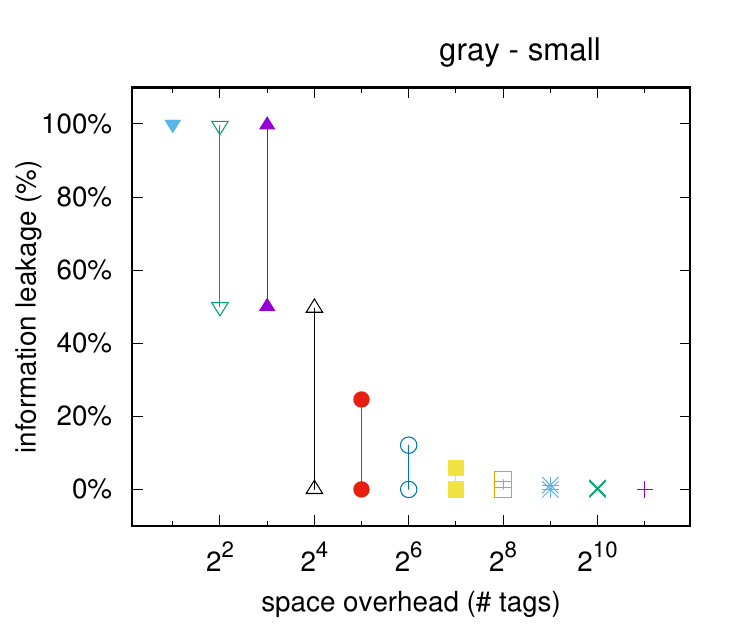}
  \end{minipage}
  \hspace{-0.3cm}%
  \begin{minipage}[t]{0.26\textwidth}
    \includegraphics[width=\textwidth, trim = {0.25cm 0.3cm 0.25cm 0}]{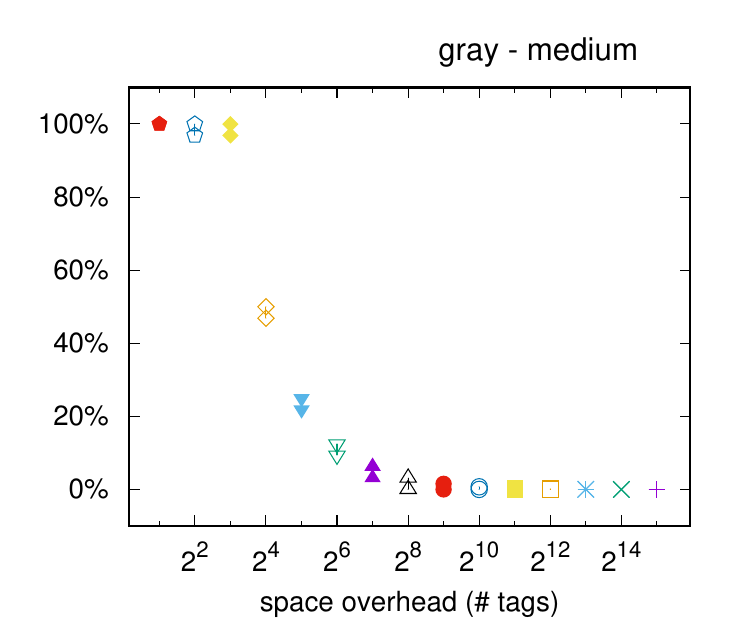}
  \end{minipage}
  \hspace{-0.3cm}%
  \begin{minipage}[t]{0.26\textwidth}
    \includegraphics[width=\textwidth, trim = {0.25cm 0.3cm 0.25cm 0}]{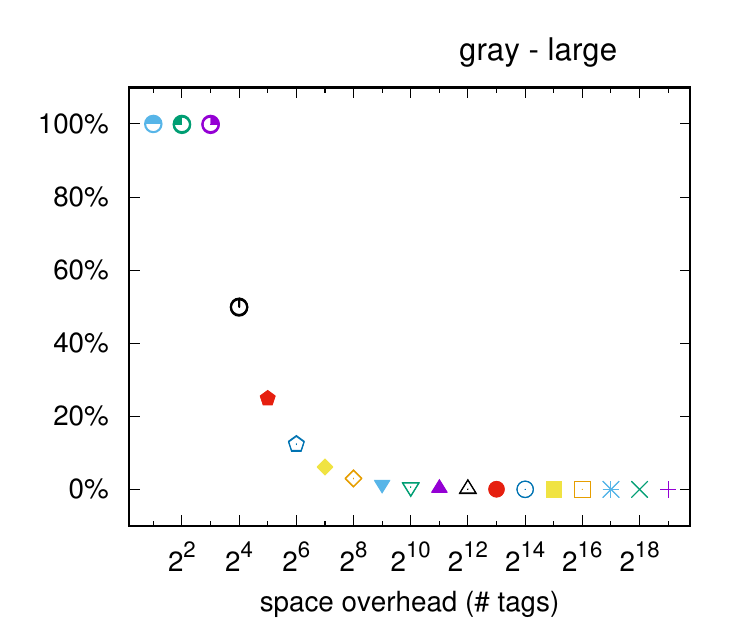}
  \end{minipage}
%
% MULTS
% 
  \centering
  \begin{minipage}[t]{0.24\textwidth}
    \includegraphics[width=\textwidth, trim = {0.95cm 0.9cm 0.35cm 0}]{plots/info_leakage/mults/mults_legend.pdf}
  \end{minipage}
  \hspace{-1.29cm}%
  \begin{minipage}[t]{0.26\textwidth}
    \includegraphics[width=\textwidth, trim = {0.25cm 0 0.25cm 0}]{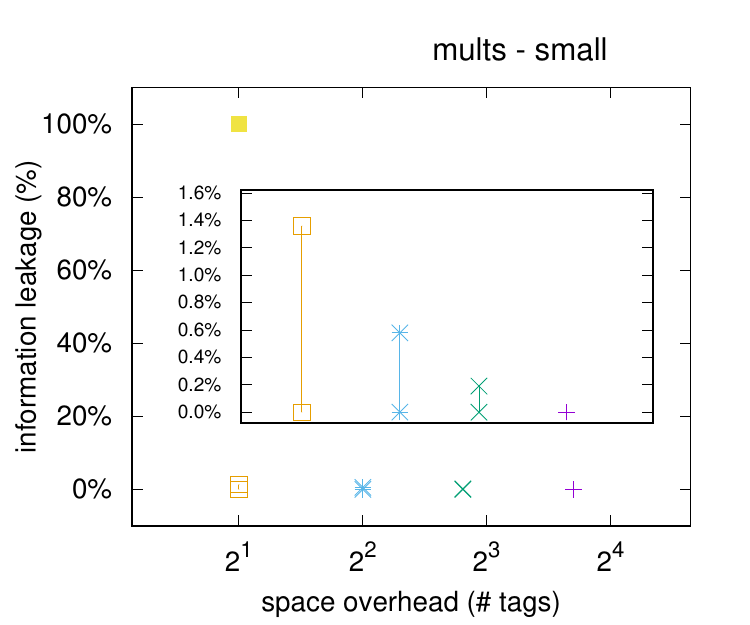}
  \end{minipage}
  \hspace{-0.3cm}%
  \begin{minipage}[t]{0.26\textwidth}
    \includegraphics[width=\textwidth, trim = {0.25cm 0 0.25cm 0}]{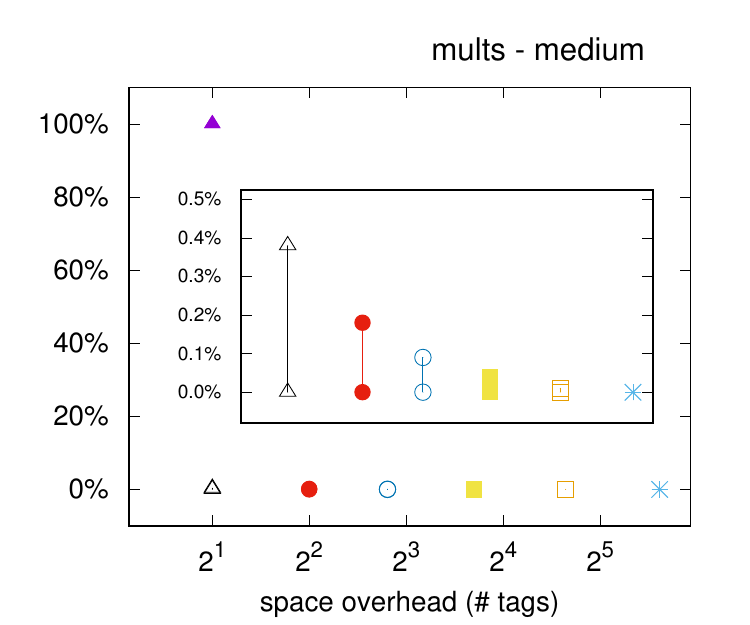}
  \end{minipage}
  \hspace{-0.3cm}%
  \begin{minipage}[t]{0.26\textwidth}
    \includegraphics[width=\textwidth, trim = {0.25cm 0 0.25cm 0}]{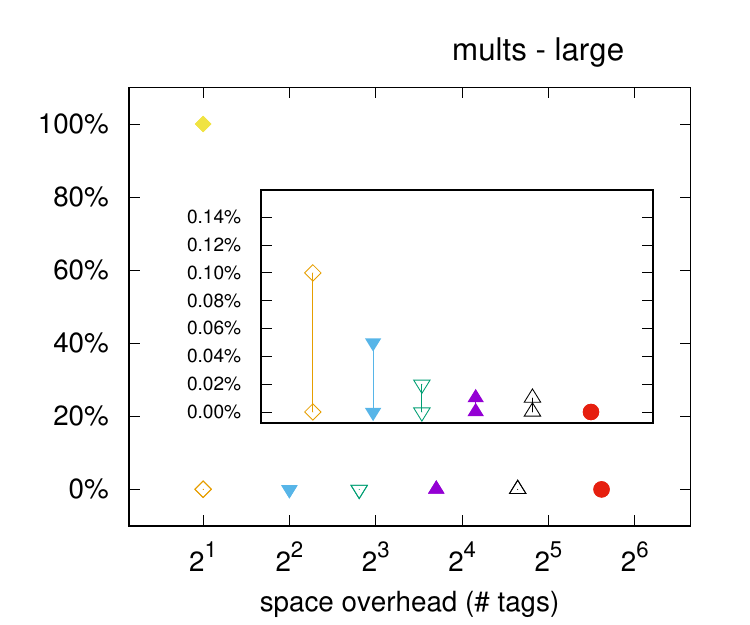}
  \end{minipage}
%
% CAPTION
%
  \caption{Measurements of space overhead for the three accelerators and 
   workloads reported in~\tablename~\ref{table:workloads}. The legend indicates
   the values of the tag offset.}\label{figure:spaceres}
  \vspace{-0.3cm}%
\end{figure*}

%
% RESULTS: execution time
%

\begin{figure*}[!tbp]
%
% MEAN 
%
 \centering
 \begin{minipage}[b]{0.26\textwidth}
    \includegraphics[width=\textwidth, trim = {0.55cm 0.85cm 0.55cm 0}]{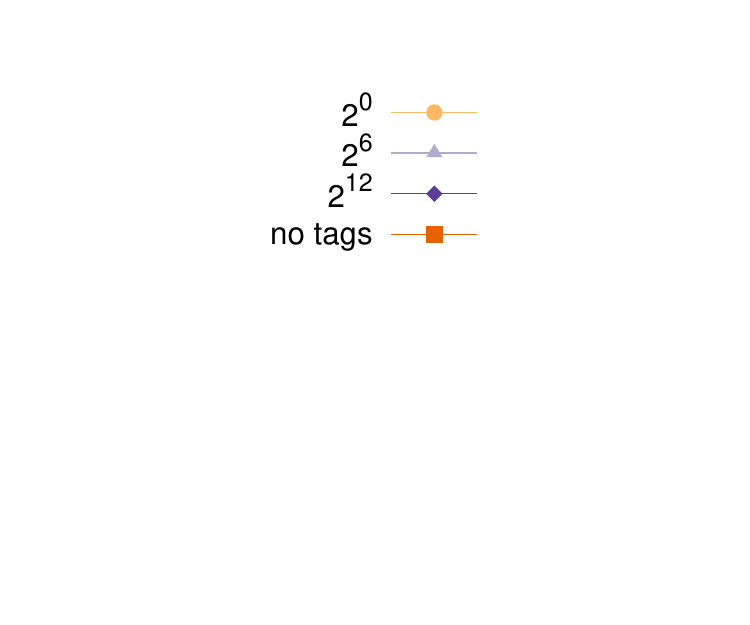}
  \end{minipage}
  \hspace{-1.38cm}%
 \begin{minipage}[b]{0.26\textwidth}
    \includegraphics[width=\textwidth, trim = {0.25cm 0.3cm 0.25cm 0}]{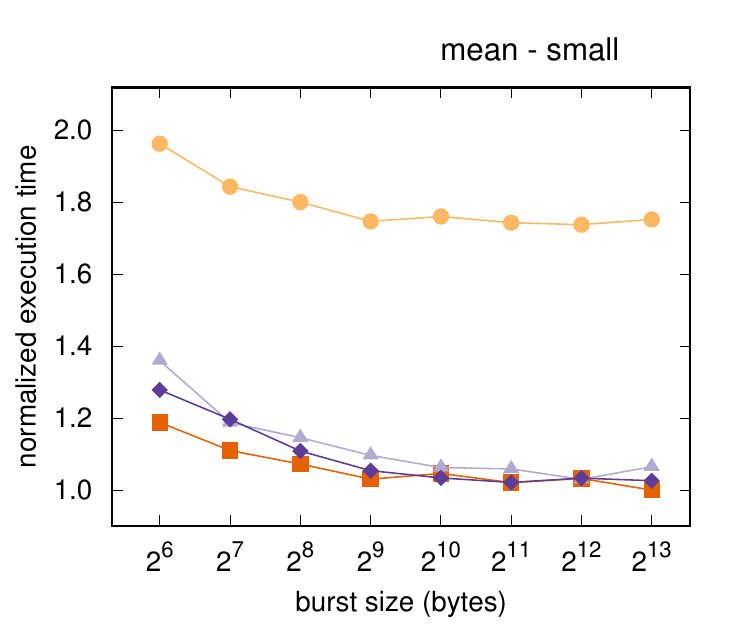}
  \end{minipage}
  \hspace{-0.38cm}%
  \begin{minipage}[b]{0.26\textwidth}
    \includegraphics[width=\textwidth, trim = {0.25cm 0.3cm 0.25cm 0}]{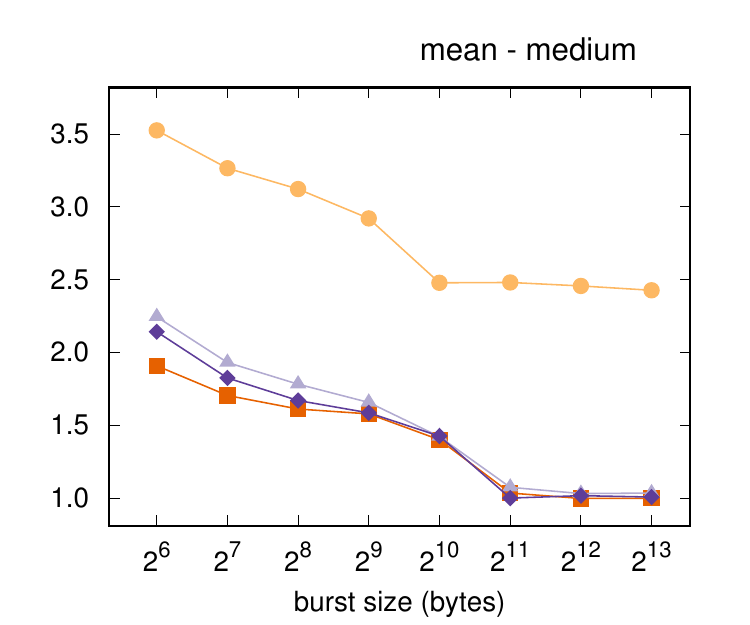}
  \end{minipage}
  \hspace{-0.38cm}%
  \begin{minipage}[b]{0.26\textwidth}
    \includegraphics[width=\textwidth, trim = {0.25cm 0.3cm 0.25cm 0}]{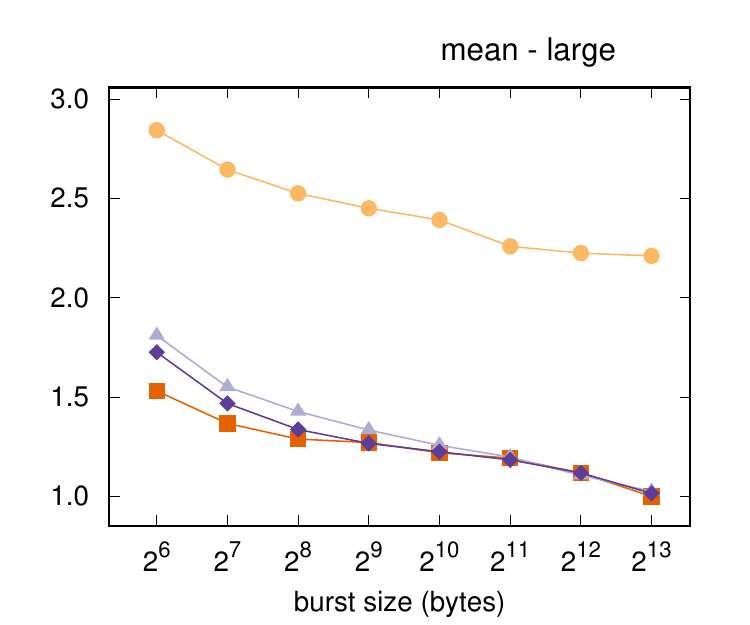}
  \end{minipage}

%
% GRAY 
%
 \centering
 \begin{minipage}[b]{0.26\textwidth}
    \includegraphics[width=\textwidth, trim = {0.55cm 0.85cm 0.55cm 0}]{plots/fpga_execution/mean/mean_legend.pdf}
  \end{minipage}
  \hspace{-1.38cm}%
 \begin{minipage}[b]{0.26\textwidth}
    \includegraphics[width=\textwidth, trim = {0.25cm 0.3cm 0.25cm 0}]{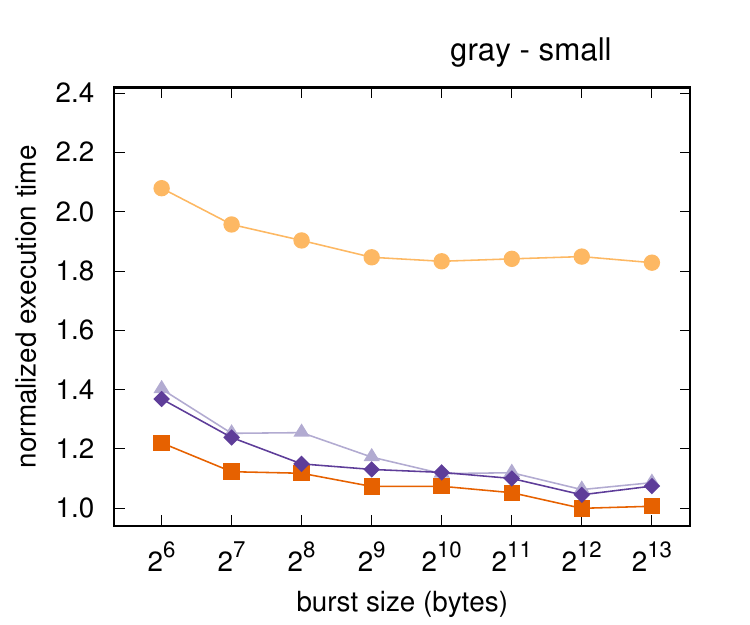}
  \end{minipage}
  \hspace{-0.38cm}%
  \begin{minipage}[b]{0.26\textwidth}
    \includegraphics[width=\textwidth, trim = {0.25cm 0.3cm 0.25cm 0}]{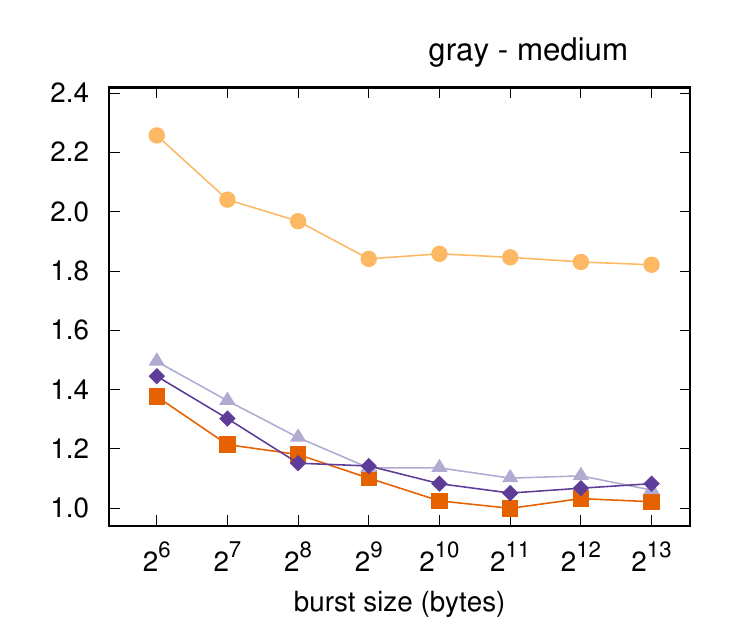}
  \end{minipage}
  \hspace{-0.38cm}%
  \begin{minipage}[b]{0.26\textwidth}
    \includegraphics[width=\textwidth, trim = {0.25cm 0.3cm 0.25cm 0}]{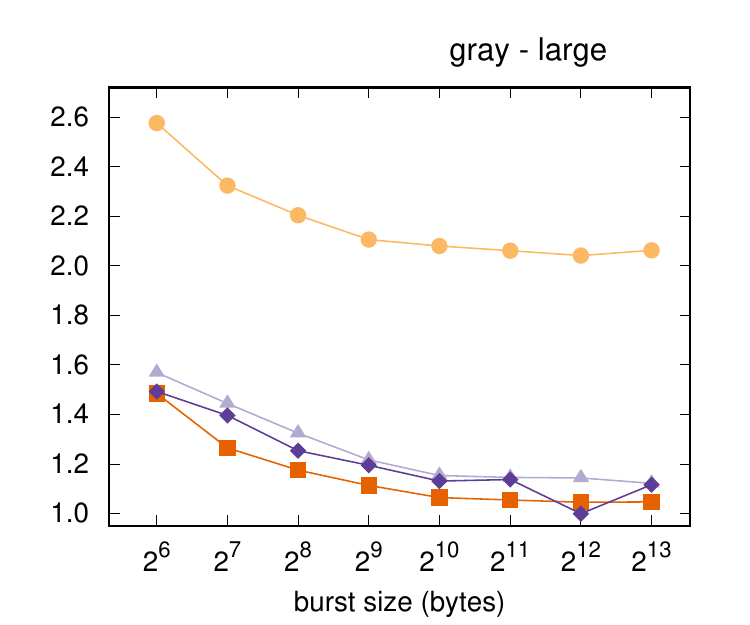}
  \end{minipage}

%
% MULTS
%
 \centering
 \begin{minipage}[b]{0.26\textwidth}
    \includegraphics[width=\textwidth, trim = {0.55cm 0.85cm 0.55cm 0}]{plots/fpga_execution/mean/mean_legend.pdf}
  \end{minipage}
  \hspace{-1.38cm}%
 \begin{minipage}[b]{0.26\textwidth}
    \includegraphics[width=\textwidth, trim = {0.25cm 0 0.25cm 0}]{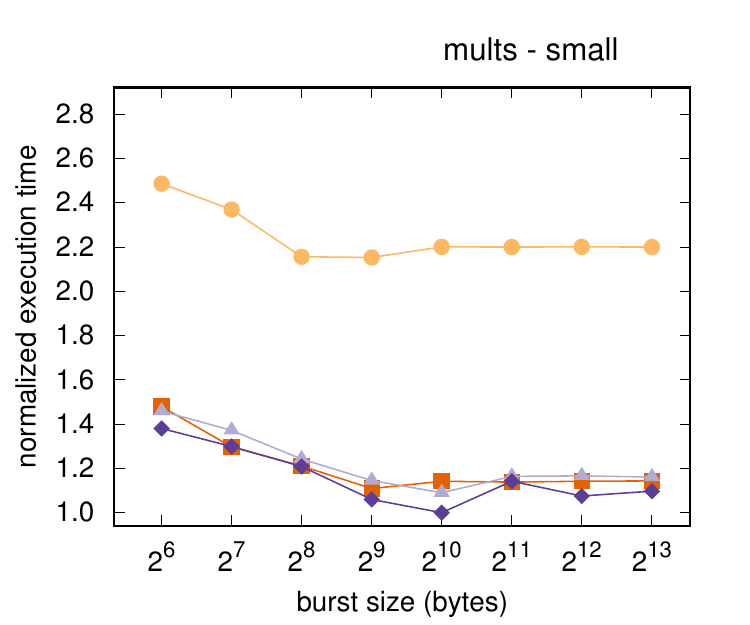}
  \end{minipage}
  \hspace{-0.38cm}%
  \begin{minipage}[b]{0.26\textwidth}
    \includegraphics[width=\textwidth, trim = {0.25cm 0 0.25cm 0}]{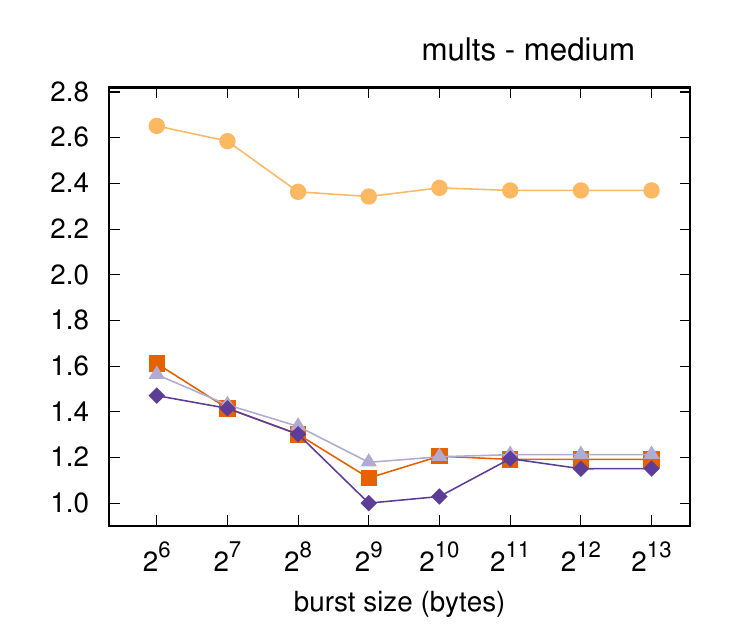}
  \end{minipage}
  \hspace{-0.38cm}%
  \begin{minipage}[b]{0.26\textwidth}
    \includegraphics[width=\textwidth, trim = {0.25cm 0 0.25cm 0}]{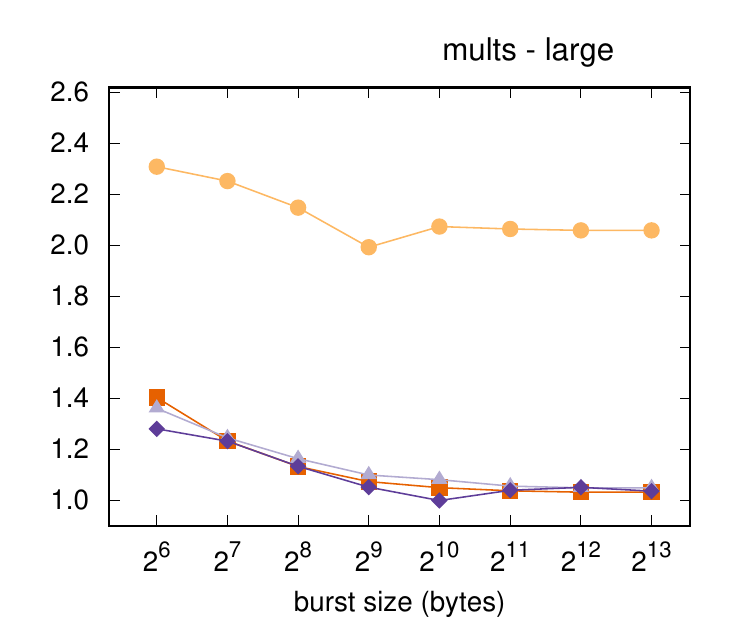}
  \end{minipage}
%
% CAPTION
%
  \caption{Measurements of execution time for the three accelerators and 
   workloads reported in~\tablename~\ref{table:workloads}. The legend indicates
   the values of the tag offset.}\label{figure:latency}
  \vspace{-0.3cm}%

\end{figure*}

%
% Experimental Evaluation
%

%
\section{Experimental Evaluation}
\label{section:results}

{
This section presents the results. We first describe the setup for the
experiments (Section~\ref{section:results:a}). Then, we discuss the results of
three design-space explorations that show how the information leakage
(\ref{section:results:b}), the space overhead (\ref{section:results:c}), and
the performance and cost of accelerators (\ref{section:results:d}) vary
depending on the accelerator, the tag density, the burst size and the workload
size.
\fillparagraph 
}

%
% WORKLOADS
%
\begin{table}[t]
\setlength{\tabcolsep}{4pt}
\caption{Workloads of the Accelerators.}\label{table:workloads}
\centering\scriptsize\begin{tabular}{cccccc}
\toprule
& {small} & {medium} & {large} \\
\midrule
\acc{MEAN}  & 128$\times$128 &
              512$\times$512 & 2048$\times$2048 \\ 
\acc{GRAY}  & 128$\times$128 &
              512$\times$512 & 2048$\times$2048 \\ 
\acc{MULTS} & 128$\times$128 &
              512$\times$512 & 2048$\times$2048 \\ 
\bottomrule
\end{tabular}
\end{table}
%

%
% Experimental Setup
%

\subsection{Experimental Setup}
\label{section:results:a}

{
We designed three accelerators (Section~\ref{section:prelim:arch}): \acc{GRAY},
\acc{MEAN}, and \acc{MULTS}. \acc{GRAY} converts a {\small RGB} image into a
grayscale image. \acc{MEAN} calculates the arithmetic mean over the columns of
a two-dimensional matrix. \acc{MULTS} performs the multiplication of a
two-dimensional matrix by its transpose. All the inputs and outputs of the
accelerators are 64-bit fixed points, except for the \acc{GRAY} input values
that are three 16-bit integer values ({\small RGB}).  We chose to implement and
analyze these accelerators since they exhibit different input/output behaviors.
In particular, \acc{GRAY} needs a single load burst to produce a store burst
since it operates in streaming.  In contrast, \acc{MEAN} and \acc{MULTS}
require multiple load bursts. \acc{MULTS} differ from \acc{MEAN} because it
needs to access the same data multiple times (at maximum two rows, or a portion
of two rows depending on the burst size, can be stored in the PLMs), and it
needs the entire input to produce few output values. We designed the accelerators in {SystemC}.
We performed high-level synthesis with Cadence Stratus HLS 17.20 and logic
synthesis with Xilinx Vivado 2017.4. We targeted a Virtex-7 {XC7V2000T} 
FPGA. We adopted the same system-level design flow for synthesizing the DIFT shell. 
%
%\fillparagraph
}

%
% Quantitative Security Analysis
%

\vspace{-0.2cm}%
\subsection{Quantitative Security Analysis}
\label{section:results:b}

{
We performed a design-space exploration of the three ac-celerators by
considering three metrics: information leakage, burst size and tag density.
For each accelerator we considered three workloads, whose characteristics are
reported in \tablename~\ref{table:workloads}. Note that the size of the
workloads determines the number of bursts of the specific accelerator.  For
example, in the case of \acc{MEAN} with burst size of $2^{7}$ bytes, $1024$
bursts are necessary to load the input matrix for the workload ``{small}''
(size: $2^{17}$ bytes). The results are reported in
\figurename~\ref{figure:leakageres}.  Each graph reports the results for a
specific accelerator and workload. The $x$-axis of each graph reports the burst
size in bytes (log scale). The $y$-axis reports the percentage of information
leakage. The colors/shapes indicate the distance between two consecutive
tags in memory, i.e., the tag offset (the larger is the tag offset, the lower
is the tag density).  We calculated the information leakage as described in
Example~\ref{example:leakage} (the worst-case scenario). For this, we did
not randomize the location of the first tag in memory to determine an upper
bound of the information leakage. For each accelerator and workload we can
identify the \textit{maximum value of tag offset that guarantees $0$\% of
information leakage}. This corresponds to the case where we interleave at least
one tag in the sequence of load bursts that are necessary to calculate a single
store burst. Note that the tag offset for all the accelerators is relatively
high compared to what we would expect for software applications, due to the
fact that the accelerators work in bursts.  From these experiments we can also
determine the \textit{minimum value of tag offset that produces $100$\% of
information leakage}. This corresponds to the case where there are no tags in
the input of the accelerator. Between the maximum and the minimum values of the
tag offset, the information leakage depends on the burst size. The larger is
the burst size, the lower is the information leakage because it is more likely
to find a tag in main memory.  The information leakage gradually decreases
by increasing the burst size until it reaches $0$\%, where the total size of the
load bursts necessary to produce a store burst has become large enough to hit a
tag in the input. By looking at the behaviors of the different accelerators we
notice that: for \acc{MEAN}, the information leakage quickly decreases since
it is a reduction operation; for \acc{GRAY}, we have more information leakage
because for one store burst we need only a single load burst; \acc{MULTS}
exhibits the lowest leakage because the algorithm needs to access most of the
input matrix to produce the first store burst. In fact, to produce the first
row of the output matrix, \acc{MULTS} needs to read all the rows of the input
matrix, i.e., the entire input. Therefore, it can leak only few output values
before realizing that a tag has been overwritten.
\fillparagraph
}

\smallskip
\textbf{Remarks}.
\textit{This experiment shows that, for any given accelerator with a certain
burst size, it is possible to determine the tag offset that guarantees a target
information leakage.} This can be determined automatically and permits to
choose the tag offset for the DIFT shell depending on the characteristics of
the specific accelerator and the workload it needs to execute.

\subsection{Space Overhead Analysis}
\label{section:results:c}
{
We performed a design-space exploration of the accelerators and workloads of
\tablename~\ref{table:workloads} by considering  three metrics: information
leakage, space overhead and tag density.  We measured the space overhead in
terms of number of tags added to the input and output of the accelerators in
main memory.  The results are reported in \figurename~\ref{figure:spaceres}.
Each graph reports the results for a specific accelerator and workload. The
$x$-axis reports the number of tags added to the input and output of the
accelerator (log scale), the $y$-axis reports the percentage of information
leakage, and the colors/shapes indicate the tag offset.  Since the information
leakage depends on the burst size (\figurename~\ref{figure:leakageres}), we
reported the information leakage for the smallest and the largest bursts
considered in Fig. 9, i.e., $2^6$ and $2^{13}$ bytes respectively. Protecting
\acc{MULTS} requires the lowest space overhead since the accelerator accesses
quickly the entire input and few tags embedded in the input are sufficient to
reduce significantly the information leakage.  \acc{MEAN} exhibits similar
space overheads because it is a reduction operation.  However, \acc{MEAN}
presents much higher information leakage due to its access pattern. \acc{GRAY}
is more difficult to protect compared to the other accelerators because it
needs a single load burst for each store burst. Thus, a higher number of tags
must be embedded in the input of the accelerator to reduce the information
leakage.  Another aspect to note is that for \acc{GRAY} and \acc{MULTS} there
is no much difference of information leakage for the smallest and the largest
bursts, while for \acc{MEAN} the burst size highly affects the information
leakage.
\fillparagraph
}

\smallskip
\textbf{Remarks}.
\textit{This experiment shows that few tags embedded in the input and
output of the accelerators are often sufficient to reduce significantly the information
leakage of accelerators.} 

\subsection{Performance and Cost Analysis}
\label{section:results:d}
\textbf{Performance}.
{ 
In order to analyze performance and cost we completed a third design-space
exploration by considering  three metrics: execution time, burst size and tag
density. For each accelerator we used the workloads in
\tablename~\ref{table:workloads}. The results are reported in
\figurename~\ref{figure:latency}.  Each graph reports the results for a
specific accelerator and workload.  The $x$-axis of each graph reports the
burst size in bytes (log scale), the $y$-axis reports the execution time
normalized to the slowest implementation, and the color/shape indicates the tag
offset.  The execution time reported in these experiments corresponds to the
time required by the accelerator to process the given workload in hardware.  To
measure the execution time for each combination of accelerator, burst size and
tag offset, we leveraged the \textit{Embedded Scalable Platforms (ESP)}
methodology~\cite{carloni2016, mantovani2016} to design an SoC that includes a
processor core (LEON3), a memory controller, and the specific accelerator.  We
ran these experiments on the FPGA by booting Linux on the processor core. The
accelerators are called through their corresponding device drivers. We
considered three values (1, 64, 4096) as tag offset  and compared the execution
time with respect to the case in which the accelerators do not use the DIFT
shell.  The graphs in \figurename~\ref{figure:latency} show that the overhead
in execution time increases as the tag offset decreases. In fact, having more
tags augments the time required by the accelerator to read the input data from
main memory and store back the results. 
%By increasing the workload size, it is
%possible to notice that the overhead of DIFT decreases, especially in the case
%of the \acc{MEAN} accelerator. This is caused by the fact that with larger
%input sizes the overhead of DIFT is partially hidden thanks to the pipelining
%of the different execution phases of the accelerator. With the largest workload
%size (the graphs in the third column)
The overhead of DIFT is relatively small for all the different workload sizes,
and we expect that with larger tag offsets it would be even smaller. 
For workload sizes much smaller than the ones reported in \tablename~\ref{table:workloads},
we expect that the overhead of DIFT would be higher since the execution times
of the accelerator would be shorter and using tags would have
a more significant impact on such executions. Note that, however, loosely coupled accelerators
typically work on relatively large data sets, as the ones used in our experiments, for which DIFT
has a low overhead.
Note that
in some graphs, the accelerators with DIFT seem to be faster than the baseline.
However, this effect is caused by the noise of the operating system running
system processes concurrently to the accelerators.
%%%
Finally, note that by increasing the burst size, the accelerators become faster
since it is more efficient performing few large bursts rather than many small
bursts~\cite{piccolboni2017b}.
\fillparagraph
}

\smallskip

{
\textbf{Cost}.
The DIFT shell is independent from the accelerators design and has a fixed
area.  For the Xilinx {XC7V2000T} FPGA, the shell requires only $\sim$$1600$
LUTs and $\sim$$1400$ flops/latches. 
\fillparagraph
}

\smallskip

\textbf{Remarks}.
{
\textit{This experiment shows that the area overhead of DIFT is negligible,
while the overhead in execution time is affected by the tag density and the
workload size.} The tag density is thus an optimization parameter for the
accelerators design: designers can strengthen or weaken the security of
accelerators in exchange of lower or higher performance, respectively. 
\fillparagraph
}

\newcommand{\riscy}{{\small RI5CY}\xspace}

\section{PULP\lowercase{ino} Case Study}
\label{section:results:pulpino}

{
To show the flexibility of \metname, we extended an open-source embedded SoC
called PULPino~\cite{gautschi2017} to {support DIFT.} We extended the \riscy
processor core in PULPino to support tagging~\cite{palmiero2018}, and we
integrated one of our accelerators. We implemented a buffer-overflow attack and
show why a holistic DIFT approach is needed to prevent software-based attacks.
While more convoluted and critical attacks can be implemented, the
software-based attack discussed here is representative of the vulnerabilities
that can be exploited in heterogeneous SoCs.
}

\begin{figure}
    \centering
    \includegraphics[width=0.47\textwidth]{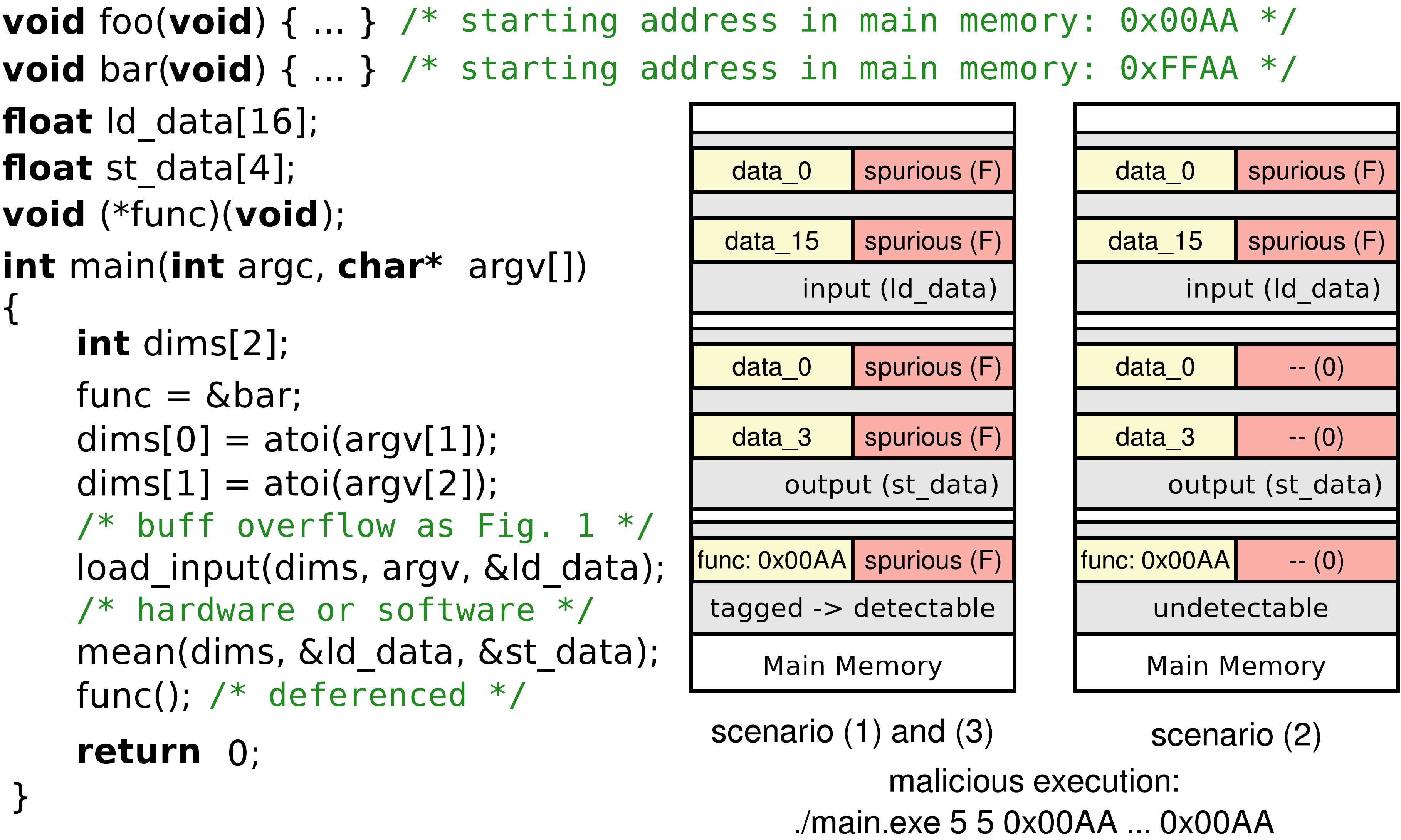}\vspace{-0.05cm}%
    \caption{Buffer-overflow attack on the PULPino SoC~\cite{gautschi2017}. In
scenario (1), {\scriptsize\texttt{mean}} is executed in software. In scenario (2) and (3),
{\scriptsize\texttt{mean}} is executed in hardware without the DIFT shell (2) and with the
DIFT shell (3).}\label{figure:pulpattack}
\vspace{-0.1cm}%
\end{figure}

\subsection{Extending PULPino with DIFT}

{
We extended \riscy, which is an in-order single-issue core with 4 pipeline
stages. We modified each stage to propagate and check the tags. We extended the
registers of \riscy as well. We added new assembly instructions for
initializing the tags stored in the register file and in the data memory.  In
this case study, we used a coupled scheme to manage the tags. We extended the
PULPino platform buses to accommodate tag transfers in parallel with the
regular data transfers.  We used four bits as width of the tag to support the
byte addressing mode of RISC-V and to distinguish only the spurious data from
the non-spurious data at the byte level.  We also integrated the \acc{MEAN}
accelerator by adapting its interface to the AXI4 interface of PULPino.  We
designed two versions of this platform.  In the first, we did not encapsulate
the \acc{MEAN} accelerator with the DIFT shell. In the second we added the DIFT
shell. We synthesized the platforms by targeting a Xilinx XC7Z020 FPGA.
\fillparagraph
}

\subsection{{Attacking PULPino with DIFT}}

{
We implemented a buffer-overflow attack on the {PULPino platform} extended with
DIFT. The code is reported in \figurename~\ref{figure:pulpattack}.  A
buffer-overflow attack occurs in the function {\small\texttt{load\_input}}.
Similarly to the attack of \figurename~\ref{figure:introattack}, the attacker
overwrites the input ({\small\texttt{ld\_data}}) with the base address of the
function {\small\texttt{foo}}.  Note that, differently from the attack of
\figurename~\ref{figure:introattack}, this attack cannot be prevented with
non-executable stack because the data structures reside in global memory. Other
attacks, such as heap overflow, can be implemented in a similar way.  The
attacker calls the function {\small\texttt{mean}} by specifying that the size
of the input is $5\times5$.  The accelerator produces the output
({\small\texttt{st\_data}}), but it also overwrites the function pointer
{\small\texttt{func}} since the output buffer can store only 4 values and not 5
{ (this causes a second buffer overflow).} The function {\small\texttt{mean}}
can be implemented in hardware or software. In both cases, we want to enforce a
policy that specifies that spurious values can never be used as pointers. We
tested the following scenarios by running the code in bare metal:
\fillparagraph
}

\begin{enumerate}
\setlength\itemsep{0.1em}

{
\item[\textbf{(1)}]
{\small\texttt{mean}} is performed in software: the buffer-overflow attack is
capable of overwriting the function pointer {\small\texttt{func}} (see the data
in main memory reported on the left); however, since the data coming from
{\small\texttt{argv}} are spurious their use as a pointer is not permitted.
Thus, an exception is raised;
}

{
\item[\textbf{(2)}]
{\small\texttt{mean}} is performed in hardware with the accelerator \acc{MEAN}
not protected with the DIFT shell: in this case the tags are not propagated
from the input to the output of the accelerator (see the data in main memory
reported on the right) and the attack is not prevented ({\small\texttt{func}}
is not tagged);
\fillparagraph
}

{
\item[\textbf{(3)}]
{\small\texttt{mean}} is performed in hardware with the accelerator
\acc{MEAN} protected by the DIFT shell: in this case the attack is prevented as
in the first case (memory reported on the left) thanks to the tag propagation
performed by the shell.
\fillparagraph
}
\end{enumerate}

%
% Discussions
%

\vspace{-0.1cm}
\section{Discussions}
\label{section:discussions}

This section discusses the benefits and limits of \metname.

%
% Coarse-grain vs. Fine-grain DIFT
%

\vspace{-0.3cm}%
\subsection{Coarse-grain Versus Fine-grain DIFT}
\label{section:discussions:granularity}

We designed the DIFT shell to extend the support of DIFT to accelerators. The
design of the accelerator is independent from the design of the shell, and the
design of the shell remains independent from the design of the accelerator.
Essentially, in our approach, the accelerator is a black box and the tags are
not propagated inside the accelerator. The shell is responsible of the tagging.
It communicates with the processor core, which decides the output tags given
the input tags (at the accelerator-level granularity). This implementation can
be called \textit{coarse-grain DIFT}, by using the same terminology currently
used for processor cores (the tags are computed at the instruction-level
granularity)~\cite{suh2004}. The alternative is \textit{fine-grain DIFT},
where the internal logic of the accelerator (or the processor) is augmented
to support tagging at the gate-level granularity, e.g.,~\cite{tiwari2009,
pilato2018}.

\smallskip

{
Both approaches have advantages and disadvantages. On one hand, fine-grain DIFT
allows a significant reduction of the false positives because it is not necessary
to take conservative choices to implement policies~\cite{pilato2018}. On the
other hand, extending the logic has a significant impact on both area and
power. This is especially true for accelerators, where up to 90\% of the area
is occupied by the PLM~\cite{lyons2012}, which needs to be extended to support
tagging. As a result, up to 31\% of additional
logic for accelerators can be necessary~\cite{pilato2018}.
Coarse-grain DIFT causes more false positives. We showed, however, that the
overhead in area is negligible and no modifications are required to the
accelerators, i.e., our approach can also be used for third-party
IPs.
\fillparagraph
}

%
% Tightly-Coupled Accelerators
%

\vspace{-0.4cm}%
\subsection{Tightly Coupled Accelerators}
\label{section:discussions:tightly}

%%%% -- Shorter version

{
In this paper we focused on loosely coupled accelerators.
Tightly coupled accelerators are required to support DIFT as well to secure
heterogeneous SoCs and avoid attacks similar to the one we implemented on
PULPino (Section~\ref{section:results:pulpino}).  Similarly to the case of
loosely coupled accelerators, two alternative implementations are possible.
With coarse-grain DIFT, the tightly coupled accelerators are black boxes and
the tags are computed at the instruction-level granularity. With fine-grain
DIFT, instead, the internal logic of the accelerators is
extended~\cite{pilato2018}. We argue that these alternatives have the same
advantages and disadvantages discussed for loosely coupled accelerators.
\fillparagraph
}

%%%% -- Longer version

%% In this paper we extended the support of DIFT to loosely-coupled
%% accelerators~\cite{cota2015}. Tightly-coupled accelerators need to support DIFT
%% as well to implement a holistic DIFT approach.  Otherwise, attacks similar to
%% the one reported in Section~\ref{section:results:pulpino} can be implemented.
%% Similarly to the case of loosely-coupled accelerators, two alternatives are
%% possible to implement DIFT on tightly-coupled accelerators.  With coarse-grain
%% DIFT, the tightly-coupled accelerator can be seen as a black box, and the
%% processor needs to decide how to propagate the tags based on the function
%% computed by the accelerator. This technique may restrict the set of policies
%% that one would want to support and lead to false positives. However, we can use
%% a low-cost and low-overhead DIFT shell, which can be designed independently
%% from the specific accelerator.  With fine-grain DIFT, instead, the internal
%% logic of the accelerator is extended similarly to the approach described
%% in~\cite{pilato2018}. This reduces the false positives, but the extension is
%% accelerator specific. It can also be prohibitively expensive for the area and
%% power overheads in which it incurs.

% Related work
%

\section{Related Work}\label{section:related}

{ 
DIFT, also called dynamic taint analysis, is a security technique to prevent
several software-based attacks~\cite{suh2004, vacha2004, chen2005, qin2006}.
Several variations of DIFT have been proposed. Most of these approaches focus
on supporting DIFT on processor cores. For example, there are approaches that
extend the processor cores and propagate the tags through the entire
architecture by extending caches, memories, and communication
channels~\cite{vacha2004, crandall2004, dalton2007, palmiero2018}.  They differ
on the target architecture, on how they manage the tags (coupled or decoupled
scheme) and on the bit widths of the tags~\cite{porquet2013}. Other approaches
adopt a co-processor to decouple the verification and the propagation of the
tags from the main processor~core~\cite{venka2008, kannan2009, deng2010}. Some
approaches are optimized for specific types of architectures, e.g., speculative
processors~\cite{chen2008}, SMT processors~\cite{ozsoy2011}, and
smarthphones~\cite{gu2013, enck2014}.  There exist also software-only
implementations of DIFT~\cite{vacha2004, qin2006, clause2007, saoji2017}, whose
overhead is usually high~\cite{newsome2005}.  Finally, there are approaches
that explore the implementation of DIFT for tag propagation at different design
abstraction levels~\cite{ardesh2017, tiwari2009} to minimize the number of
false positives.  All these approaches are complementary to \metname. In fact,
\metname can be used to easily extend the support for DIFT, implemented on
processors, to accelerators.
}

\smallskip

{ 
Most of the approaches on hardware-based DIFT focus on supporting DIFT on
processor cores rather than entire SoCs.  To the best of our knowledge, there
are only two works in the literature in the direction of a holistic DIFT
implementation. {\small WHISK}~\cite{porquet2013} targets SoCs with
loosely coupled accelerators. {\small WHISK} implements fine-grain DIFT on
accelerators, differently from \metname that realizes a coarse-grain DIFT
approach.  \metname interleaves the tags with the data, while in {\small WHISK}
the tags are stored in a different region of memory. Finally, while we define
the concept of information leakage which is an accelerator-dependent metric, in
{\small WHISK} the authors used the concept of security proportionality, which
is the amount of tags supplied as input to the accelerator.  Another work
related to accelerators is {TaintHLS}~\cite{pilato2018}, which is a methodology
to automatically add support for fine-grain DIFT on accelerators developed with
high-level synthesis. TaintHLS cannot be used to secure hard IP cores as
well as soft IP cores designed at RTL without licensable high-level descriptions.
Also, by using a fine-grain approach, TaintHLS can incur in significant area
overhead (up to 31\%) because the accelerators logic must be extended. 
\fillparagraph
}

%
% Concluding Remarks
%

\section{Concluding Remarks}
\label{section:remarks}

{
We presented \metname, a flexible methodology to design a circuit shell that
extends DIFT to loosely coupled accelerators. The design of the DIFT shell is
independent from the design of the accelerators and vice versa. This allows
designers to quickly support DIFT on their accelerators in heterogeneous SoCs.
We studied the effect of the shell on the cost and performance of the accelerators
by running experiments on a FPGA. We showed that the cost of the shell is
negligible compared to the cost of the accelerators. The performance overhead
depends on the tag density, which is a parameter that can be tuned by designers
to strengthen or weaken the security guarantees of the particular accelerator. To
quantitatively measure such security guarantees, we defined a metric,
called information leakage.  We performed a multi-objective design-space
exploration and showed that we can synthesize implementations of accelerators
encapsulated with the DIFT shell that present different trade-offs in terms of
performance, cost and information leakage.  We also showed that, for any given
accelerator, it is possible to determine the minimum amount of tagging for DIFT that
guarantees absence of information leakage.  Finally, we presented a case study
where we extended PULPino to support DIFT and we showed the effectiveness of
the DIFT shell in preventing a buffer-overflow attack that exploits a
loosely coupled accelerator.
}

%
% Acknowledgment 
%

\section*{Acknowledgment}

The authors would like to thank the anonymous reviewers for their valuable
comments and helpful suggestions. The authors would like also to thank Simha
Sethumadhavan and Vasileios Kemerlis for their valuable feedback and Paolo
Mantovani for the support with the experimental framework. This work was
supported in part by DARPA SSITH (HR0011-18-C-0017).

\vspace{-0.15cm}%

\bibliographystyle{IEEEtran}
\bibliography{IEEEabrv,refs}

%
% Biographies
%

\vskip -2.5\baselineskip plus -1fil

\begin{IEEEbiography}[{
\includegraphics[width=0.9in,clip,keepaspectratio]{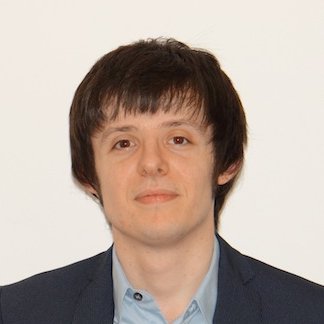}}] {Luca
Piccolboni} (S'15) received the B.S. degree ({\em summa cum laude}) in Computer
Science from the University of Verona, Verona, Italy, in 2013, and the M.S.
degree in Computer Science and Engineering ({\em summa cum laude}) from the
University of Verona in 2015. He is currently working toward the Ph.D. degree
in Computer Science at Columbia University, New York, NY, USA.  His research
interests include design and verification of embedded systems, with particular
regard to computer-aided design, high-level synthesis, hardware acceleration,
and system security.
\end{IEEEbiography}

\vskip -2.4\baselineskip plus -1fil

\begin{IEEEbiography}
[{\includegraphics[width=1in,height=1.25in,clip,keepaspectratio]{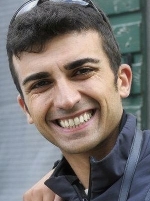}}]
{Giuseppe Di Guglielmo} (S'06, M'09) received the Laurea degree ({\em summa cum laude}) in
Computer Science from the University of Verona, Verona, Italy, in 2005, and the
Ph.D. degree in Computer Science from {the University} of Verona in 2009. He is
currently an Associate Research Scientist with the Department of Computer
Science, Columbia University, New York, NY, USA. He has authored over 50
publications. His current research interests include system-level design and
validation of system-on-chip platforms. In this context, he collaborated in
several US, Japanese and Italian projects. He is a member of the IEEE.
\end{IEEEbiography}

\vskip -2.1\baselineskip plus -1fil

\begin{IEEEbiography}[{
\includegraphics[width=1in,height=1.25in,clip,keepaspectratio]{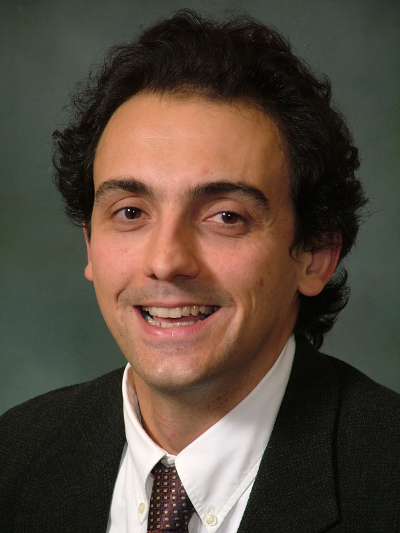}}]
{Luca P. Carloni} (S'95, M'04, SM'09, F'17) received the Laurea degree 
(\textit{summa cum laude}) in electrical engineering from the Universit\`{a} di
Bologna, Bologna, Italy, in 1995, and the M.S. and Ph.D. degrees in electrical
engineering and computer sciences from the University of California at
Berkeley, Berkeley, CA, USA, in 1997 and 2004, respectively. He is Professor of
Computer Science at Columbia University in the City of New York, NY, USA. He
has authored over 130 publications and holds two patents. His current research
interests include system-on-chip platforms, system-level design, distributed
embedded systems, and high-performance computer systems.  Dr. Carloni was a
recipient of the Demetri Angelakos Memorial Achievement Award in 2002, the
Faculty Early Career Development (CAREER) Award from the National Science
Foundation in 2006, the ONR Young Investigator Award in 2010, and the IEEE CEDA
Early Career Award in 2012. He was selected as an Alfred P. Sloan Research
fellow in 2008. His 1999 paper on the latency-insensitive design methodology
was selected for the Best of ICCAD, a collection of the best papers published
in the first 20 years of the IEEE International Conference on Computer-Aided
Design. In 2013, he served as the General Chair of Embedded Systems Week, the
premier event covering all aspects of embedded systems and software. He is a
Senior Member of the Association for Computing Machinery.
\end{IEEEbiography}

\end{document}